\documentclass[12pt]{article}
 \pdfoutput=1
 \usepackage{graphicx}
  \usepackage{hyperref}
  \usepackage{epsfig}
  \usepackage{amsmath}
  \usepackage{amssymb}
  \usepackage{color}
  \usepackage[nosort]{cite}
  \newlength{\abstractwidth}
  \newcommand{\be}{\begin{equation}}
  \newcommand{\ee}{\end{equation}}
  
  \renewcommand{\title}[1]{\vbox{\center\bf{\Large{#1}}}\vspace{5mm}}
  \renewcommand{\author}[1]{\vbox{\center#1}\vspace{5mm}}
  \newcommand{\address}[1]{\vbox{\center\em#1}}
  \newcommand{\email}[1]{\vbox{\center\tt#1}\vspace{5mm}}

\begin{document}

\begin{titlepage}
\begin{center}
\hfill \\
\hfill \\
\vskip 1cm

\title{Stringy effects in scrambling}

\author{Stephen H. Shenker${}^1$ and Douglas Stanford${}^{1,2}$}

\address{
 ${}^1$ Stanford Institute for Theoretical Physics {\it and} \\
 Department of Physics, Stanford University \\
 Stanford, CA, USA\\
 ${}^2$ School of Natural Sciences, Institute for Advanced Study \\
 Princeton, NJ, USA
}

\email{sshenker@stanford.edu,
stanford@ias.edu}

\end{center}
  
  \begin{abstract}
  In \cite{SS} we gave a precise holographic calculation of chaos at the scrambling time scale.   We studied the influence of a small perturbation, long in the past, on a two-sided  correlation function in the thermofield double state.   A similar analysis applies to squared commutators and other out-of-time-order one-sided correlators \cite{Shenker:2013yza, RSS, kitaev}.  The essential bulk physics is a high energy scattering problem near the horizon of an AdS black hole. The above papers used Einstein gravity to study this problem; in the present paper we consider stringy and Planckian corrections. Elastic stringy corrections play an important role, effectively weakening and smearing out the development of chaos. We discuss their signature  in the boundary field theory, commenting on the extension to weak coupling.  Inelastic effects, although important for the evolution of the state, leave a parametrically small imprint on the correlators that we study. We briefly discuss ways to diagnose these small corrections, and we propose another correlator where inelastic effects are order one. 
  \end{abstract}

  \end{titlepage}

\tableofcontents

\baselineskip=17.63pt

\section{Introduction}
The usual way to think about high energy effects in string theory, or more generally in quantum gravity,  is to envision high energy scattering experiments.   The deviation of string theory amplitudes from pointlike particle amplitudes at large center of mass energy, $\sqrt{s}$, signals the novel extended nature of these degrees of freedom.   One characteristic  effect is the transverse spreading of strings.  In a high energy collision, the characteristic transverse size  grows like $\ell_s \sqrt{\log s \ell_s^2}$, where $\ell_s$ is the string length.

Black holes  provide a laboratory where such high energy processes are important even when no explicitly high energy quanta are injected into the system.   The basic reason for this was first realized in the work of Hawking and Unruh on quantum radiation from black holes.   Near a black hole horizon, outside static (Schwarzschild) observers are accelerating  relative to a global reference frame (like Kruskal) and their time represents the rapidity of a boost. Energies measured by Schwarzschild observers in frames separated by time $t$ differ by a factor of $e^{2\pi t/\beta}$, where $\beta$ is the inverse Hawking temperature.  This exponential ratio can produce energies of string scale, Planck scale or even higher,  opening up the question of stringy or transplanckian effects in black holes.

The quantum consequences of these large boosts have been the object of extensive study.  Susskind \cite{Susskind:1993aa} investigated the transverse and longitudinal spreading of strings and its relation to the stretched horizon.   The transverse part was interpreted as branched diffusion in \cite{Mezhlumian:1994pe}.   The potential importance of the  characteristic time when this boost becomes Planckian, $t_* \sim \beta\log m_p \beta$,  was pointed out in  \cite{Schoutens:1993hu, Susskind:1993aa,Mezhlumian:1994pe}. This timescale made another appearance in the work of Hayden and Preskill \cite{Hayden:2007cs}, who made the connection between the quantum chaotic process of fast scrambling \cite{dankert:2009a,emerson:2005a,harrow:2009a,arnaud:2008a,brown:2010a,diniz:2011a,Brown:2012gy,Lashkari:2011yi} and the Planckian boost time $t_*$.   Sekino and Susskind \cite{Sekino:2008he} further connected these concepts to gauge/gravity duality.

The firewall proposal \cite{Almheiri:2012rt,Almheiri:2013hfa,Marolf:2013dba} has re-emphasized the importance of these effects, as transplanckian physics provides an obstruction \cite{obstruction} to building the interior using the pullback-pushforward technique \cite{Freivogel:2004rd}.  Other recent work on the effect of large boosts in related contexts includes discussions of brane dynamics\cite{Sila}, of stretched string production\cite{Silverstein:2014yza}, and of longitudinal spreading\cite{longitudinal}.

Building on work of van Raamsdonk and collaborators \cite{VanRaamsdonk}  we \cite{SS} investigated scrambling in the two sided eternal black hole in AdS/CFT.   We gave a sharp holographic derivation of the butterfly effect, where large boosts again play the central role.  Our analysis began with the thermofield double state (TFD) of two CFTs `$L$' and `$R$', dual to the eternal AdS Schwarzschild black hole \cite{israel, Maldacena:2001kr}. This state has a large degree of special two-sided correlation, diagnosed e.g. by two-sided correlation functions $\langle \phi_L \phi_R\rangle$. We perturbed this state by applying an operator $W$ at time $t$ on the left boundary.\footnote{In \cite{SS} we referred to $t$ as $t_w$.} This operator adds a small amount of energy to the system, of order one thermal quantum. However, if $t$ is sufficiently large, the perturbation powerfully disrupts the two-sided correlation. The bulk explanation for this is that $W$ produces a shock wave whose energy in the global $t=0$ frame appears boosted to $\beta^{-1}e^{2\pi t/\beta}$ where $\beta$ is the inverse Hawking temperature.  When this energy becomes of order the mass scale of the black hole, i.e., $t \sim t_* = \frac{\beta }{2 \pi} \log S$ (where $S$ is the entropy per thermal volume in the CFT), the effect of the shock on $\phi$ propagation becomes important and substantial decorrelation of left and right degrees of freedom occurs.

Because of the large boost at scrambling time, the energies involved in this process are very high. In our original work \cite{SS}   we studied this system using Einstein gravity and gave a preliminary discussion of the potentially important string and planck scale corrections.   The basic point of the present paper is to give a more systematic analysis of these corrections. We will show that the important physics for the computation of the correlator, even at large $t$, is scattering at energies with $G_N s \sim 1$ in AdS units. At these scales, inelastic effects are parametrically subleading. However, elastic stringy corrections are important. String theory leads to two related corrections to the picture of scrambling. First, as might be expected from \cite{Susskind:1993aa}, stringy effects smear out the region of decorrelation over a scale $\rho \sim \ell_s \sqrt{t}/\ell_{AdS}$. Second, due to Regge-ization, the scattering amplitude grows more slowly with $s$ than in pure gravity. This leads to a ``string corrected'' scrambling time
\be
t_* = \frac{\beta}{2\pi}\left[1 + \frac{d(d-1)\ell_s^2}{4\ell_{AdS}^2}+...\right]\log S\label{corrected-scrambling}
\ee
where $d$ is the space-time dimension of the boundary theory.

\subsection{Correlation functions that probe chaos}
Before getting started, we would like to place the problem of computing $\langle \phi_L\phi_R\rangle_W$ in a slightly more general context, and to introduce some new notation. (The discussion here has some overlap with \cite{virasoro}.)
\begin{figure}
\begin{center}
\includegraphics[scale = .8]{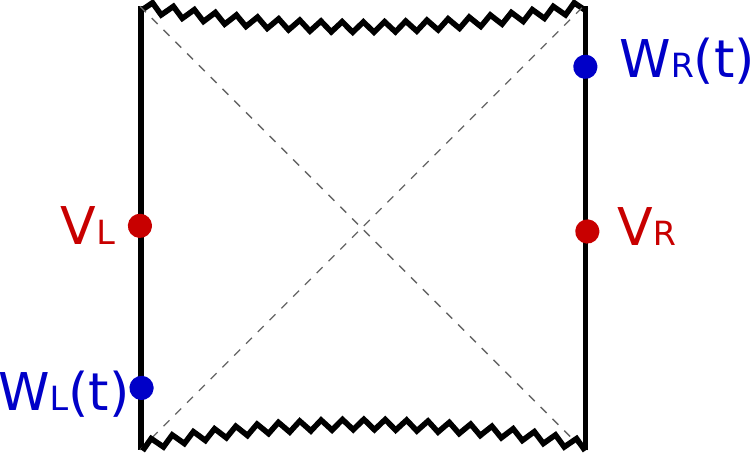}
\caption{Locations on the Penrose diagram of the various operators described in the text.}\label{fig1}
\end{center}
\end{figure}
The case we originally considered, in \cite{SS}, was a two-sided correlation function of the form
\be
\langle W_R(t) V_L V_R W_R(t)\rangle = \langle V_L W_R(t) V_R W_R(t)\rangle \label{WVWV1}
\ee
where the expectation value indicates the thermofield double state, and we have switched $L\leftrightarrow R$ compared to \cite{SS}.\footnote{A word on conventions. Given an operator $V$ in a single copy of the CFT, we define $V_R = 1 \otimes V$, acting on the $R$ system. We define $V_L = V^T\otimes 1$, where the transpose is taken in the energy eigenbasis. Under the Killing time evolution, $V_L(t) = e^{-iH_Lt}V_Le^{iH_Lt}$ and $V_R(t) = e^{iH_Rt}V_Re^{-iH_Rt}$.} In this paper, we will take $V$ and $W$ to be approximately local operators, that raise the energy of the thermal state by an amount of order $\beta^{-1}$. (For simplicity of notation, we will also assume the operators are Hermitian and that they have vanishing one-point functions.) In a suitably chaotic system, we expect correlation functions of this type to become small at large $t$, regardless of the specific choice of $V,W$. This is supported by the analysis of \cite{SS}.

Other correlation functions can also be used to probe the same physics. In particular, we can consider completely one-sided thermal correlators, like
\be
\langle V\,W(t)V\, W(t)\rangle.\label{WVWV2}
\ee
Here and in the remainder of the paper, operators with omitted subscripts are assumed to act on the $R$ system, and operators without time arguments are assumed to be at $t =0$. Other one-sided orderings, such as $VVW(t)W(t)$ and $VW(t)W(t)V$ are not directly sensitive to chaos. At large time, they approach $\langle VV\rangle \langle WW\rangle$. This implies that the behavior of the squared commutator $\langle [W(t),V]^2\rangle$ is determined by (\ref{WVWV2}). The fact that this correlator becomes small at late times indicates that the squared commutator becomes large, of order $\langle VV\rangle\langle WW\rangle$.

Correlation functions similar to (\ref{WVWV2}) were analyzed by Larkin and Ovchinnikov \cite{larkin} for a single particle in a chaotic potential. More recently, the order-one commutator between any two operators $V,W$ was connected to the butterfly effect in \cite{Almheiri:2013hfa}. The commutator was studied using holography in \cite{Shenker:2013yza, RSS}, and in \cite{RSS} as a diagnostic of the growth with time of local operators. Kitaev \cite{kitaev}, has also examined correlation functions such as (\ref{WVWV2}). Building on \cite{larkin} he made the connection between the initial exponential behavior of these correlation functions and Lyapunov exponents, also calculated the correlators using gravitational shock wave scattering in a theory dual to gravity, and pointed  out the unusual quantization of these Lyapunov exponents in such a theory.  Similar correlation functions have also been studied in large $c$ 2d CFTs \cite{virasoro, Jackson:2014nla}.

Finally, we can also consider a configuration with two operators on each side: 
\be
\langle V_LW_R(t)V_RW_L(t)\rangle.\label{WVWV3}
\ee
This arrangement is closely related to the work of \cite{Cornalba:2006xk,Cornalba:2006xm,Cornalba:2007zb} on high energy scattering in vacuum AdS. Although the bulk physics in these references is rather similar to what we will study, the boundary interpretation is quite different. In \cite{Cornalba:2006xk,Cornalba:2006xm,Cornalba:2007zb}, the operators $V,W$ are explicitly high energy. In our case, they are thermal scale; the high energies are a result of the boosting effect of time evolution in the black hole geometry.

\begin{figure}[ht]
\begin{center}
\includegraphics[scale = .8]{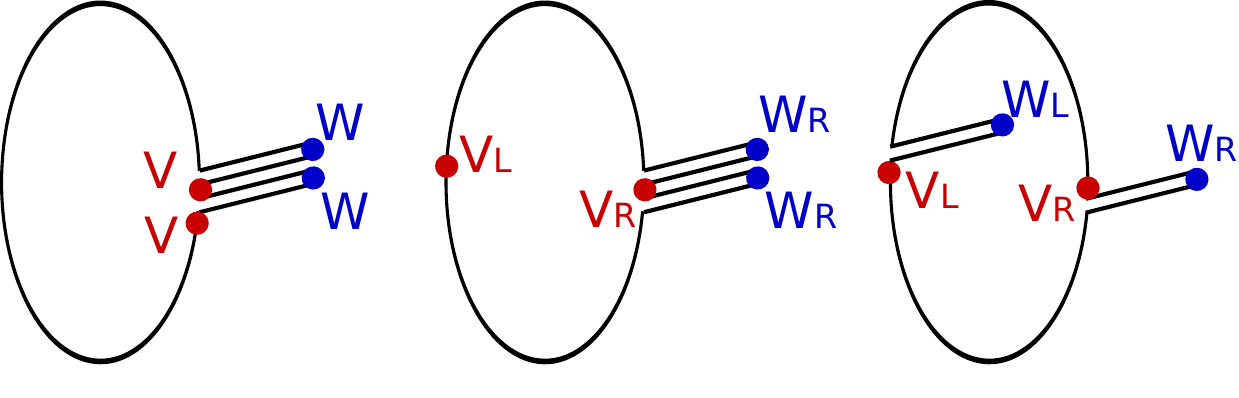}
\caption{The path integral contours that define (\ref{WVWV2}), (\ref{WVWV1}), and (\ref{WVWV3}), respectively. The circle is the periodic imaginary time direction, and the folds represent the real-time evolution to produce the $W(t)$ operators. The contour ordering is the same in each case, so the correlators are related by adding or subtracting imaginary time $\beta/2$ to one or two of the operators.}\label{contours}
\end{center}
\end{figure}

Although we have described these correlation functions separately, we emphasize that they are all related by analytic continuation. For concreteness, we will often focus on the purely one-sided configuration (\ref{WVWV2}). The first function (\ref{WVWV1}) can be obtained by adding $-i\beta/2$ to the time argument of the first $V$ operator. To get the third (\ref{WVWV3}), we also add $+i\beta/2$ to the time of the second $W$ operator. This continuation is explained in Fig.~\ref{contours}. All of these correlation functions can be obtained by analytically continuing the Euclidean correlator to a second sheet. See \cite{virasoro} for a recent explanation of the necessary continuation.

\subsection{Outline}
Equations (\ref{WVWV1}), (\ref{WVWV2}), and (\ref{WVWV3}) define three related observables that are sensitive to the same basic chaos. In \S~\ref{skinematics}, we will see that their holographic computation is controlled by the same bulk physics -- a high-energy scattering problem near the bifurcation surface of a black hole. The different correlation functions translate to somewhat different wave functions to be folded against the same scattering amplitude. High energy scattering in gravity and in string theory has been studied extensively, and our analysis of these correlation functions will consist of adapting flat space results to the weakly curved spacetime region near the bifurcation surface. We will do this analysis in three stages.

First, in \S~\ref{sone}, we will review and clarify the calculation of the above correlation functions using the elastic eikonal approximation in gravity \cite{'tHooft:1987rb,Amati:1987uf,Verlinde:1991iu,Kabat:1992tb,Cornalba:2007zb,Brower:2007qh}. This generalizes our original calculation from \cite{SS}, which was done using a geodesic estimate of correlations in the background of a spherically symmetric shock wave.

Next, in \S~\ref{stringy}, we will address tree-level stringy corrections to the gravitational calculation, using the techniques of \cite{Brower:2006ea}. Bulk stringy effects are related to finite coupling in the dual CFT, so it is natural to expect that these tend to weaken the effect of chaos. Indeed, we find that they delay the decrease of the correlation functions (\ref{WVWV1}), (\ref{WVWV2}), and (\ref{WVWV3}), as indicated by the longer scrambling time in (\ref{corrected-scrambling}). Also, the transverse spreading of strings discussed above has the effect of smearing out the region in which the correlation function is affected.

Finally, in \S~\ref{stwo}, we will assess inelastic effects, following flat space intuition from \cite{Amati:1987uf}. Surprisingly, these have very little effect on the correlation functions described above, despite the fact that large time $t$ translates to an exponentially large relative boost of the $W$ and $V$ operators. For sufficiently large $t$, this boost is enormous, and the scattering problem is likely to be dominated by extremely inelastic processes. The key point is that the correlation functions (\ref{WVWV1}), (\ref{WVWV2}), and (\ref{WVWV3}) involve an integral over scattering momenta, weighted by wave functions. The oscillating phase from the elastic eikonal amplitude tends to suppress contributions from large momenta. Even at very large $t$, the important part of the integral is over modes for which the center of mass energy in the collision satisfies $G_N s \sim 1$ in AdS units.\footnote{More precisely, the important region of integration is where the tree level amplitude is order one. For impact parameters of order AdS scale, this is equivalent to $G_N s\sim 1$.} At this energy, inelastic effects are parametrically subleading.

Of course, as $t$ increases, these modes becomes farther out in the tails of the wavefunctions, and most of the state $V\,W(t)|TFD\rangle$ becomes dominated by inelastic phenomena. Still, when we form the correlator (\ref{WVWV2}) by contracting with $\langle TFD|V\,W(t)$, only the $G_N s\sim 1$ tails are important. This gives us good control over the correlation functions. However, a disappointing corollary is that these correlators do not expose truly Planckian physics, limiting our ability to connect high energy bulk physics to boundary chaos.

In the Discussion, we will describe other observables that might probe inelasticities more directly. We also suggest how scrambling might be visible in perturbation theory, by computing (\ref{corrected-scrambling}) in a small $\lambda$ expansion.

Appendix \ref{brownian} contains an analysis of scrambling in a completely connected two-local spin system evolving with time-dependent random couplings. This system has a $\log n$ scrambling time \cite{Lashkari:2011yi}. We compute a correlation function of the type studied using holography in this paper, finding good qualitative agreement. 

\section{Kinematics}\label{skinematics}
 The holographic calculation of each of the correlation functions (\ref{WVWV1}), (\ref{WVWV2}), and (\ref{WVWV3}) consists of sewing wave functions together with a high energy scattering amplitude. The problem separates into a kinematical piece, which depends on the particular choice of correlation function, and a dynamical piece -- the amplitude itself. In this section, we will study the kinematics. 

The setting of the scattering problem is an AdS black hole geometry. It will be useful to describe the metric in Kruskal coordinates,
\be
ds^2 = -a(uv)dudv + r^2(uv)dx^idx^i.
\ee
Here, $r(uv)$ is the standard Schwarzschild $r$ coordinate. The Schwarzschild time $t$ is also a function of $u$ and $v$. The bulk space-time dimension is $D = d+1$, so that $i$ runs over $d-1$ values. The two components of the horizon are at $u =0$ and $v = 0$; they meet at the bifurcation surface $u = v = 0$. We will use
\be
a_0 = a(0), \hspace{20pt} r_0 = r(0)
\ee to indicate the values at the horizon. In writing this metric, we have assumed a planar black hole, but a similar analysis applies to large black holes with spherical horizons. 

The result of this section is a formula for the correlation function
\be
D(\{t_i,x_i\}) = \langle V_{x_1}(t_1)W_{x_2}(t_2)V_{x_3}(t_3)W_{x_4}(t_4)\rangle\label{correlator-to-compute}
\ee
where all operators act on the right boundary, as in (\ref{WVWV2}), and we have made their spatial locations explicit. We are interested in a configuration of times with $(x_1,t_1)\approx (x_3,t_3)$ and $(x_2,t_2)\approx (x_4,t_4)$ but $t_1-t_2\gg \beta$. We will focus on the fully one-sided configuration here, but we emphasize that the other configurations (\ref{WVWV1}) and (\ref{WVWV3}) can be obtained by the half-period continuations discussed above. They can also be obtained directly, as we indicate near the end of the section.

We will first state the formula, then explain it, then derive it. The formula is
\be
D(\{t_i,x_i\}) = \frac{a_0^4}{(4\pi)^2}\int e^{i\delta(s,|x-x'|)}  \Big[p_1^u \psi_1^*(p_1^u,x)\psi_3(p_1^u,x)\Big]\Big[p_2^v\,\psi_2^*(p_2^v,x')\psi_4(p_2^v,x')\Big]\label{overlapformula}
\ee
where the integral runs over transverse positions $x,x'$ with measure factors $r_0^{d-1}$, and over null momenta $p_1^u,p_2^v$. The factor $e^{i\delta(s,b)}$ is a two-to-two scattering amplitude, defined as a function of transverse separation $b$ and the Mandelstam-like variable
\be
s = a_0\,p_1^up_2^v.\label{mandelstams}
\ee
The wave functions are Fourier transforms of bulk-to-boundary propagators\footnote{Here, and for the remainder of the paper, we assume the $V,W$ operators are single-trace. For multi-trace operators the wave functions are more complicated but a similar construction applies.} along either the $u = 0$ or $v = 0$ horizons:
\begin{align}
\psi_1(p^u,x) &= \int dv \,e^{ia_0p^u v/2}\,\langle \phi_v(u,v,x) V_{x_1}(t_1)^\dagger\rangle|_{u=0}\label{psi1} \\
\psi_2(p^v,x) &= \int du \,e^{ia_0p^v u/2}\,\langle \phi_w(u,v,x) W_{x_2}(t_2)^\dagger\rangle|_{v=0}\label{psi2} \\
\psi_3(p^u,x) &= \int dv \,e^{ia_0p^u v/2}\,\langle \phi_v(u,v,x) V_{x_3}(t_3)\rangle|_{u=0} \label{psi3}\\
\psi_4(p^v,x) &= \int du \,e^{ia_0p^v u/2}\,\langle \phi_w(u,v,x) W_{x_4}(t_4)\rangle|_{v=0} \label{psi4}
\end{align}
In this expression, $\phi_v,\phi_w$ are the bulk fields dual to $V,W$, and the CFT operators $V,W$ are represented via the `extrapolate' dictionary as limits of bulk operators near the boundary. As with all formulas in this paper, there is no implicit time ordering; correlation functions are ordered as written. Finally, note that we have $\dagger$'s acting on two of the operators. This is important, despite our assumption that the operators are Hermitian, because we will sometimes consider complex values of the time parameters.

\subsection{Derivation}
Although the correlation function $D(\{t_i,x_i\})$ is a one-sided quantity, depending only on the density matrix of the $R$ system, we will find it useful to think about the correlation function in the two-sided purification provided by the thermofield double state $|TFD\rangle$. Our starting point, as in \cite{SS,RSS}, is to represent $D$ as an overlap of two states
\be
|\Psi\rangle = W(t_2)^\dagger V(t_1)^\dagger|TFD\rangle, \hspace{20pt}|\Psi'\rangle = V(t_3)W(t_4)|TFD\rangle.
\ee
Here and below, we suppress the position subscripts. Both of these states contain two quanta, (or, more generally, two sets of quanta), created by the $W$ and $V$ operators. If the difference in times $t_2-t_1$ is large, the relative boost between the quanta is also large $\sim e^{\frac{2\pi}{\beta}(t_2-t_1)}$. In a symmetric frame, the quanta created by the $W$ operator will have large $p^v$, and will be traveling close to the $u = 0$ horizon. The quanta created by the $V$ operator will have large $p^u$ and will be traveling near $v = 0$.

\begin{figure}[h]
\begin{center}
\includegraphics[scale = .7]{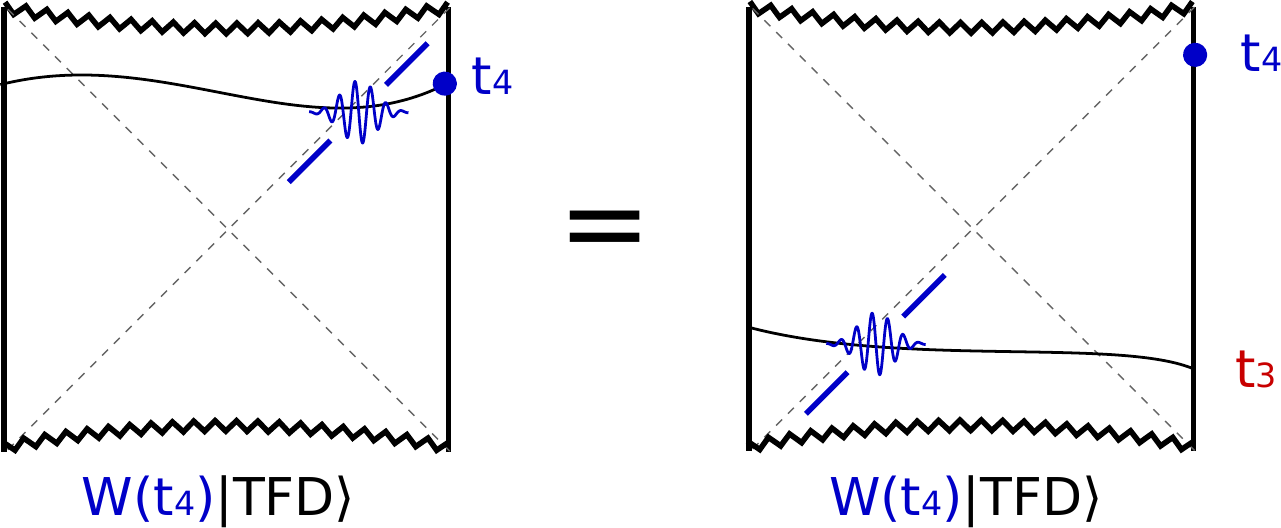}
\caption{The one-particle state $W(t_4)|TFD\rangle$ can be represented on any bulk slice.}\label{slices}
\end{center}
\end{figure}

\begin{figure}[ht]
\begin{center}
\includegraphics[scale = .7]{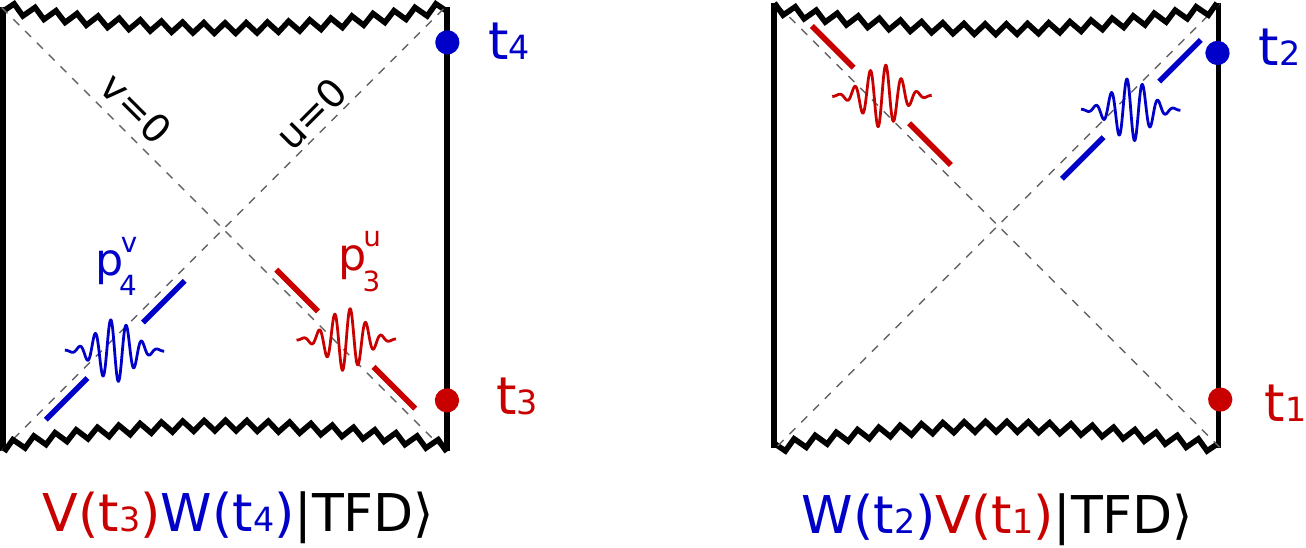}
\caption{The correlation function (\ref{correlator-to-compute}) is an inner product of these two states. As explained in the text, changing the ordering of these operators changes an `in' state to an `out' state.}\label{scatter2}
\end{center}
\end{figure}

The important difference between $|\Psi\rangle$ and $|\Psi'\rangle$ is whether the quanta are created `above' or `below' their eventual collision. Or, more precisely, the difference is whether the states are `in' or `out' states with respect to a global notion of time that increases upwards. In order to understand this point, we will follow the procedure from \cite{Shenker:2013yza}. This involves building the state $V(t_3)W(t_4)|TFD\rangle$ in two steps. First, we act with the operator $W(t_4)$ on $|TFD\rangle$. This creates a one-particle state, similar to the one shown in the left panel of Fig.~\ref{slices}.
We can represent this one-particle state on different bulk slices. In order to act with the $V(t_3)$ operator, it is convenient to evolve the state backwards to an early slice that touches the $R$ boundary at time $t_3$, as shown in the right panel. On this slice, the $W$ and $V$ quanta are spacelike related, so it is simple to act with the $V$ operator. The result is an `in' state, as shown in the left panel of Fig.~\ref{scatter2}.

This `in' state can be described using Klein-Gordon wave functions. These are simply bulk to boundary propagators from the relevant points on the boundary. In high-energy scattering processes, it is often useful to describe wave functions in terms of longitudinal momentum and transverse position. We would therefore like to decompose these Klein Gordon wave functions in a basis of $(p^v,x)$ for the $W$ particle and $(p^u,x)$ for the $V$ particle. In a curved background, the notion of momentum is not unique, so in order to be precise, we will proceed as follows. First, we represent the $W$ state in the Hilbert space on the $v = 0$ surface, and the $V$ state on the $u = 0$ slice. We then Fourier transform in the remaining null coordinates, getting the wave functions $\psi_3$ and $\psi_4$ from Eq (\ref{psi3}) and (\ref{psi4}).
Given these wave functions, the `in' state is a two-particle state
\be
 V(t_3)W(t_4)|TFD\rangle = \int  \psi_3(p_3^u,x_3)\psi_4(p_4^v,x_4)\,|p_3^u,x_3; p_4^v,x_4\rangle_{in},
\ee 
where the integral runs over all exposed variables. The symbol $|p_3^u,x_3; p_4^v,x_4\rangle$ represents a tensor product of a ket vector $|p_3^u,x_3\rangle$ in the Hilbert space at $u = 0$, and a ket vector $|p_4^v,x_4\rangle$ in the Hilbert space at $v = 0$. The normalization is
\be
\langle p^v,x|q^v,y\rangle = \frac{a_0^2\,p^v}{4\pi r_0^{d-1}}\delta(p^v-q^v)\delta^{d-1}(x-y)
\ee
and a related expression for the $|p^u,x\rangle$ vectors. Working through a similar procedure, one finds the other state is an `out' state,
\be
W(t_2)V(t_1)|TFD\rangle = \int \psi_1(p_1^u,x_1)\psi_2(p_2^v,x_2)\,|p_1^u,x_1; p_2^v,x_2\rangle_{out},
\ee
where $\psi_{1,2}$ are defined in Eq (\ref{psi1}) and (\ref{psi2}).\footnote{One minor subtlety that we have glossed over is that the operators must be smeared out in time, in order for the correlation functions to be finite. We will assume that this smearing is over a thermal scale $\beta$. An alternative is to assign small imaginary times. These should be positive for the $3,4$ operators and negative for the $1,2$ operators.}

We now take the overlap of the states $|\Psi\rangle$ and $|\Psi'\rangle$:
\be
D = \int \psi_3^*(p_3^u,x_3)\psi_4^*(p_4^v,x_4)\psi_1(p_1^u,x_1)\psi_2(p_2^v,x_2) \ {}_{out}\langle p_3^u,x_3;p_4^v,x_4|p_1^u,x_1;p_2^v,x_2\rangle_{in},\label{overlap}
\ee
where the integral runs over all displayed variables. If the relative boost $e^{\frac{2\pi}{\beta}t}$ is large, and we work in the center of mass frame, the wave functions will prefer a region of integration where the momenta $p_1^u,p_2^v,p_3^u,p_4^v$ are all large. In a scattering problem with these kinematics, the other null momenta $p_1^v,p_2^u,p_3^v,p_4^u$ will be small, and momentum conservation (which holds up to scales set by the curvature) implies that $p_1^u\approx p_3^u$ and $p_2^v\approx p_4^v$. High energy also implies a small time interval for the scattering, so the $x$ coordinates are also approximately conserved. The amplitude is then essentially diagonal in the $p_1^u,p_2^v,x_1,x_2$ variables. In other words, we can approximate
\be
|p_1^u,x_1; p_2^v,x_2\rangle_{out} \approx e^{i\delta(s,b)}|p_1^u,x_1; p_2^v,x_2\rangle_{in} + |\chi\rangle\label{diag}
\ee
where $|\chi\rangle$ represents the inelastic component of the scattering; it is orthogonal to all `in' states that consist of a single $W$ particle and a single $V$ particle. The content of this equation is that, within the two-particle subspace, the scattering matrix is simply multiplication by a complex number $e^{i\delta}$. The (complex) function $\delta$ is the dynamical input to the correlation function; it will be the focus of the remainder of the paper. Using boost invariance and translation invariance along the transverse space, we have expressed $\delta$ as a function of $s$ and $b$, where
\be
b =|x_1-x_2|
\ee is the impact parameter, and $s$ is defined in Eq.~(\ref{mandelstams}) above. Subtituting (\ref{diag}) in (\ref{overlap}), we get the desired result (\ref{overlapformula}).\footnote{In this derivation, we have neglected contributions to the amplitude in which a quantum created by $V$ is annihilated by $W$ and vice versa. Such contributions will be small after a few thermal timescales.}

\subsection{A two-sided case}
We emphasized above that formulas for the other correlation functions (\ref{WVWV1}) and (\ref{WVWV3}) can be obtained by analytic continuation of the formula (\ref{overlapformula}). However, we will also give a brief sketch of the direct construction in the totally two-sided case (\ref{WVWV3}). As before, we view the correlation function as an overlap of two states:
\be
|\Psi\rangle = V_R(t_3)W_L(t_4)|TFD\rangle \hspace{20pt} |\Psi'\rangle = W_R(t_2)^\dagger V_L(t_1)^\dagger |TFD\rangle.
\ee
Here, identifying the above as `in' and `out' states is actually somewhat easier than in the case treated above. The reason is that the operators in each of the states $|\Psi\rangle$ and $|\Psi'\rangle$ are already spacelike related, so the intermediate step of evolving the quanta back to an early slice, or forward to a late slice, is unnecessary. 

\begin{figure}[ht]
\begin{center}
\includegraphics[scale = .7]{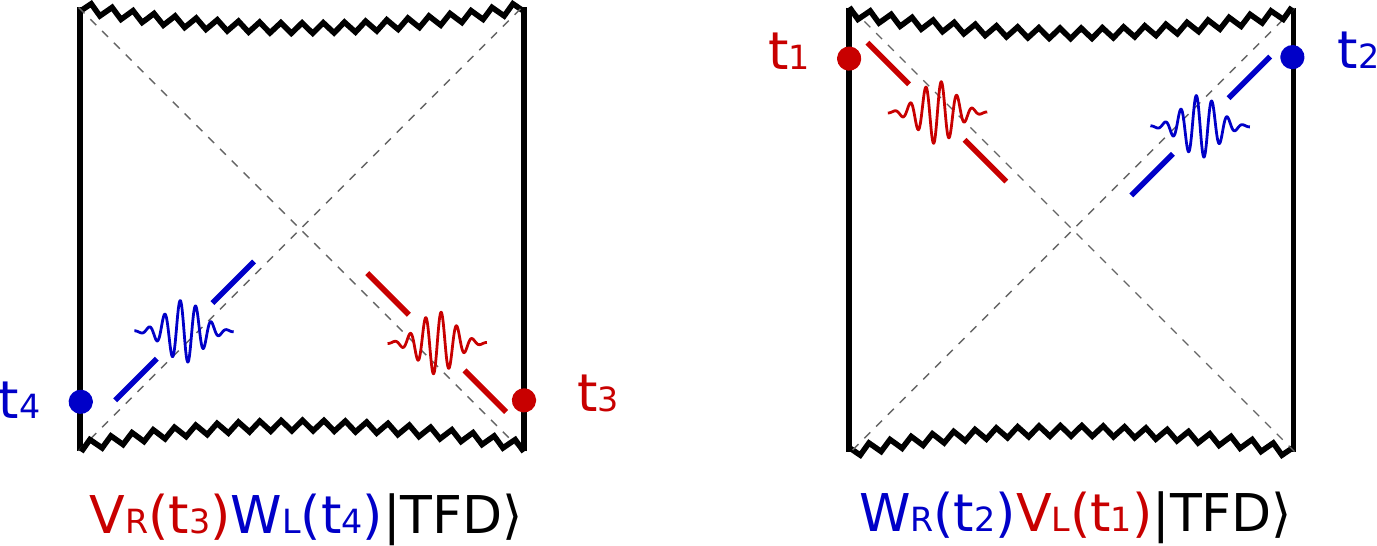}
\caption{The `in' and `out' states in the completely two-sided case (\ref{WVWV3}).}\label{scatter1}
\end{center}
\end{figure}

The two states are illustrated in Fig.~\ref{scatter1}. It is clear that $|\Psi\rangle$ is an `in' state and $|\Psi'\rangle$ is an `out' state. Their overlap is again an integral of the form (\ref{overlapformula}). The wave functions $\psi_2$ and $\psi_3$ are exactly as in (\ref{psi2}) and (\ref{psi3}), and the wave functions $\psi_1$ and $\psi_4$ are given by
\begin{align}
\psi_1(p^u,x) &= \int dv \,e^{ia_0p^u v/2}\,\langle \phi_v(u,v,x) V_{L,x_1}(t_1)^\dagger\rangle|_{u=0}\label{psi12} \\
\psi_4(p^v,x) &= \int du \,e^{ia_0p^v u/2}\,\langle \phi_w(u,v,x) W_{L,x_4}(t_4)\rangle|_{v=0} \label{psi42}.
\end{align}

\section{Elastic eikonal gravity approximation}\label{sone}
\subsection{The basic amplitude}
In the previous section, we showed how to calculate correlation functions (\ref{WVWV1}), (\ref{WVWV2}), and (\ref{WVWV3}) as a wave function overlap (\ref{overlapformula}), weighted by a scattering amplitude $e^{i\delta(s,b)}$. In this section, we will study the amplitude using the elastic eikonal approximation in gravity. At fixed impact parameter $b$ in a purely gravitational theory, this approximation is valid for small $G_N$, with $G_N s$ held fixed. 

The flat space eikonal approximation has been studied by a number of authors, including \cite{'tHooft:1987rb,Amati:1987uf,Verlinde:1991iu,Kabat:1992tb}. It has also been studied  in pure AdS by \cite{Cornalba:2007zb,Brower:2007qh}. Our analysis in the AdS black hole setting will be very closely related to the flat space black hole analysis by 't Hooft \cite{'tHooft:1990fr} and Kiem, Verlinde, and Verlinde \cite{Kiem:1995iy}. We will find the path integral perspective of Kabat and Ortiz \cite{Kabat:1992tb} particularly convenient. This approach breaks the approximation into two parts: {\it (i)} linearizing the gravity Lagrangian (diagramatically, this is equivalent to restricting to crossed ladder diagrams) and {\it (ii)} treating the scattering particles as fixed stress energy sources, following their unscattered classical trajectories (this is related to the simplifed propagators normally used in the ladder diagrams). The amplitude $e^{i\delta}$ is then given by a Gaussian integral over the metric. This can be done by evaluating the action, expanded to quadratic order in the metric, on the classical solution sourced by the particle trajectories:  
\be
\delta(s,b) = S_{cl}.
\ee

The relevant classical solution is simply the sum of the Coulomb fields of the two particles. In general, finding Coulomb fields in a curved background is difficult, but for high energy particles close to the $u = 0$ and $v = 0$ horizons, the field takes a simple shock wave form. We will focus on the field sourced by the $W$ particle moving along the $u = 0$ horizon at transverse position $x_2$. For large $p_2^v$ and correspondingly sharp localization in the $u$ direction, the stress tensor is
\be
T_{uu} = \frac{a_0}{2r_0^{d-1}} p_2^v \delta(u)\delta^{d-1}(x-x_2).
\ee
Here we are using an exact $\delta$ function in the transverse space; in the physical problem that we will study, wave functions have finite width in the $x$ space of order $\ell_{AdS}$. As long as the separation $|x_{12}|$ is larger than this width, we can use the simple $\delta$ function form.

The Coulomb field associated to this stress energy was worked out in \cite{SS,RSS}, following Aichelburg and Sexl \cite{Aichelburg:1970dh}, Dray and 't Hooft \cite{Dray:1984ha} and Sfetsos \cite{Sfetsos:1994xa}. The metric is a shock wave localized on the $u = 0$ horizon,
\be
ds^2 = -a(uv)dudv + r^2(uv)dx^idx^i + h_{uu}du^2,\label{shockmetric}
\ee
where
\be
h_{uu}(u,v,x) = \frac{8\pi G_N a_0}{r_0^{d-3}}\,p_2^v\,\delta(u) f(x-x_2).\label{huu}
\ee
The transverse profile satisfies
\begin{align}
(-\partial_x^2 + \mu^2)f(x) = \delta^{d-1}(x),\label{transverseprofile}
\end{align}
and behaves at large $\mu |x|$ as 
\be
f(|x|) = \frac{\mu^{\frac{d-4}{2}}}{2(2\pi |x|)^{\frac{d-2}{2}}}e^{-\mu |x|}.
\ee
The quantity $\mu^2$ will play an important role in this paper. It is always positive, and is given in terms of the inverse temperature $\beta$ and the horizon radius $r_0$ as
\be
\mu^2 = \frac{2\pi(d-1)r_0}{\beta},\label{mu}
\ee
where we assume dimensionless $x$ coordiantes and dimensionful $r,t$. For a planar AdS black hole we have
\be
\frac{\mu^2}{r_0^2} = \frac{d(d-1)}{2\ell_{AdS}^2}.\label{mu2}
\ee

There is a similar solution $h_{vv}\propto p^u_1\delta(v)f(x-x_1)$ sourced by the $V$ particle on the $v = 0$ horizon. The action is a sum of three terms
\be
\frac{1}{2}\int d^{d+1}x\sqrt{-g} \Big[ h_{uu}\mathcal{D}^2h_{vv} + h_{uu}T^{uu} + h_{vv}T^{vv}\Big]\label{action1}
\ee
where $\mathcal{D}^2$ is a differential operator, and $g$ refers to the unperturbed metric. On the classical solution, the last two terms are equal, and the first is equal in magnitude but opposite in sign. Therefore
\begin{align}
S_{cl} &= \frac{1}{2}\int d^{d+1}x \sqrt{-g} \  h_{uu} T^{uu} \\
&= \frac{4\pi G_N}{r_0^{d-3}}s \,f(x_{12}).
\end{align}
We conclude that, in the elastic eikonal approximation,
\be
\delta(s,b) = \frac{4\pi G_N}{r_0^{d-3}}s \,f(b).\label{gravity}
\ee

\subsection{Relation to free propagation on a shock background}
This amplitude has a very nice interpretation in terms of wave function overlaps on a shock wave background \cite{'tHooft:1987rb,Kabat:1992tb}. In this section, we will explain this interpretation, and make contact with our original calculation in \cite{SS}.

\begin{figure}
\begin{center}
\includegraphics[scale = .8]{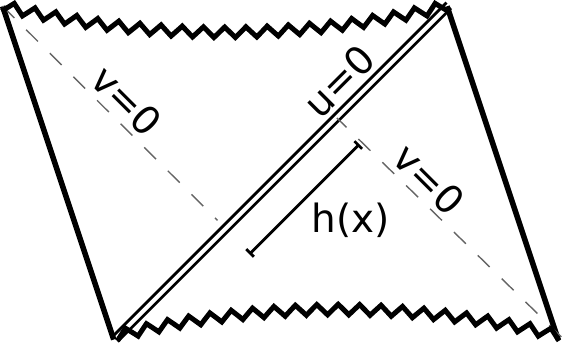}
\caption{The geometry with a shock wave at $u = 0$. Notice that the $v = 0$ surface is discontinuous by amount $h(x)$.}\label{shockpicture}
\end{center}
\end{figure}

To begin, consider a geometry with a shock wave on the $u = 0$ horizon, shown in Fig.~\ref{shockpicture}.
 The metric is of the form (\ref{shockmetric}). We will parametrize $h_{uu}$ as
\be
h_{uu} = a_0h(x)\delta(u).\label{shockgeom}
\ee
The geometry can be understood as two halves of the unperturbed black hole, glued together at $u=0$ with a shift in the $v$ direction of size
\be
\delta v(x) = h(x).\label{shift}
\ee

On this background, we consider an overlap of two states, one ($\Psi$) created by an operator to the left of the shock, and the other ($\Phi$) created by an operator to the right. Let us suppose that these states are represented, in the unperturbed background, by Klein Gordon wave functions $\Psi(u,v,x)$ and $\Phi(u,v,x)$. We can take the overlap on any bulk slice. Following Ref.~\cite{Hofman:2008ar}, we choose to evaluate it on a null slice $u = \epsilon$, just to the left of the shock. The overlap of Klein Gordon wave functions is
\be 
\langle \Psi|\Phi\rangle = 2ir_0^{d-1}\int dv dx\, \tilde{\Psi}(u,v,x)^*|_{u=\epsilon}\,\partial_v\tilde{\Phi}(u,v,x)|_{u=\epsilon}.
\ee
where $\tilde{\Psi}$ and $\tilde{\Phi}$ represent the wave function in the background that includes the shock. On the slice $u = \epsilon$ these are simply related to $\Psi,\Phi$. Using (\ref{shift}), the wave functions are just $\tilde{\Psi} = \Psi$ and $\tilde{\Phi}(\epsilon,v,x) = \Phi(\epsilon,v-h(x),x)$. Fourier transforming in $v$ on the $u = 0$ slice, we can write the resulting overlap as
\be\label{intcorr}
\langle \Psi|\Phi\rangle = \frac{a_0^2r_0^{d-1}}{4\pi}\int dp^u dx\, e^{ia_0\,p^u\, h(x)/2}\Big[p^u \Psi^*(p^u,x)\Phi(p^u,x)\Big].
\ee
The factor $a_0/2$ appears in the exponential because $v$ is conjugate to a lower-index momentum $p_v<0$, which is related to the upper index $p^u$ by a metric factor, $p_v = -a_0p^u/2$.

Now, we return to our scattering problem by comparing this overlap to the integral (\ref{overlapformula}) with the elastic eikonal gravity amplitude (\ref{gravity}) inserted. This is proportional to \footnote{A very similar equation has been independently derived by Kitaev\cite{kitaev}.}
\be
\int \exp\Big[\frac{4\pi i G_N a_0}{r_0^{d-3}}p_1^up_2^v \,f(b)\Big]  \Big[p_1^u \psi_3^*(p_1^u,x_1)\psi_1(p_1^u,x_1)\Big]\Big[p_2^v\,\psi_4^*(p_2^v,x_2)\psi_2(p_2^v,x_2)\Big].\label{amp1}
\ee
In order to compare with (\ref{intcorr}), we write the phase factor as $ia_0\,p_1^u\, h(x_2)/2$ with 
\be\label{hx}
h(x) = \frac{8\pi i G_N}{r_0^{d-3}} \,p_2^v\, f(|x -x_1|).
\ee
Using (\ref{shockgeom}) and (\ref{huu}), we see that this is the null shift associated to the shock generated by a mode of momentum $p_2^v$ traveling on the $u = 0$ horizon. The overlap (\ref{amp1}) can therefore be thought of as an overlap of the $\psi_{1,3}$ wave functions for the $V$ quantum in a geometry created by the $\psi_{2,4}$ quanta ($W$). Because the $\psi_{2,4}$ states involve a superposition of different momenta, we have to integrate over the strength of the shock, weighted by the momentum factors. Of course, the expression is totally symmetric in interchanging $(1,3)\leftrightarrow (2,4)$, so we can also think about it as an overlap of the $\psi_{2,4}$ wave functions in a shock wave background sourced by a superposition of $\psi_{1,3}$ quanta. The reason for this somewhat unusual `either but not both' interpretation can be traced to the fact that, on a classical solution, the action reduced to just one of the two equal final terms in (\ref{action1}).

In \cite{SS}, we studied a correlation function of the type $\langle W_L V_L V_R W_L\rangle$ by putting in a classical shock wave geometry for the $W$ operators and computing the correlation of the $\langle V_L V_R\rangle$ operators in that fixed background (using a geodesic estimate). In other words, we restricted to the average momentum $p_2^v$ in the state created by the $W$ operator. This is a good approximation in an asymmetric setup where the $W$ operator has a larger dimension $\Delta$  than the $V$ operator (for example, it might source more quanta, as we imagined in \cite{SS}), and the wave functions are therefore more sharply peaked in $p_2^v$. In general, one has to integrate over both $p_1^u$ and $p_2^v$.

\subsection{The integral over momenta}
In this section we will consider the integral over momenta and transverse position in (\ref{overlapformula}), given the form for the phase $\delta$ in (\ref{gravity}). 

\subsubsection*{AdS${}_3$ example}
As a warmup, we will start by considering the example of Rindler AdS${}_3$, which can be written in Kruskal coordinates as (in AdS units)
\be
ds^2 = -\frac{4 du dv}{(1+uv)^2} + \frac{(1-uv)^2}{(1+uv)^2}dx^2.
\ee
This can be understood as the dual to a spatially infinite 2d CFT, at temperature $\beta = 2\pi$. However, since the geometry is a piece of pure AdS${}_3$, the bulk-to-boundary propagator is known exactly:
\be
\langle \phi(u,v,x) \mathcal{O}(t_1,x_1)\rangle =\frac{c_\mathcal{O}}{(u e^{t_1} -ve^{-t_1} + \cosh (x-x_1))^\Delta}.
\ee
Fourier transforming along the respective horizons, we find the wave functions (\ref{psi1}-\ref{psi4})
\begin{align}
\psi_1(p^u,x) &= \theta(p^u)\frac{-2\pi i\,c_V\,e^{t_1^*}}{\Gamma(\Delta_V)} \left(-2ip^u e^{t_1^*}\right)^{\Delta_V-1}e^{2ip^u\,e^{t_1^*}\cosh (x-x_1)}\\
\psi_2(p^v,x') &= \theta(p^v)\frac{2\pi i\, c_W\,e^{-t_2^*}}{\Gamma(\Delta_W)} \left(2ip^ve^{-t_2^*}\right)^{\Delta_W-1}e^{-2ip^v\,e^{-t_2^*}\cosh (x'-x_2)} \\
\psi_3(p^u,x) &= \theta(p^u)\frac{-2\pi i\,c_V\,e^{t_3}}{\Gamma(\Delta_V)} \left(-2ip^u e^{t_3}\right)^{\Delta_V-1}e^{2ip^u\,e^{t_3}\cosh (x-x_1)}\\
\psi_4(p^v,x') &= \theta(p^v)\frac{2\pi i\, c_W\,e^{-t_4}}{\Gamma(\Delta_W)} \left(2ip^ve^{-t_4}\right)^{\Delta_W-1}e^{-2ip^v\,e^{-t_4}\cosh (x'-x_2)}.
\end{align}
These formulas are appropriate for operators on the (complexified) right boundary.

To compute the correlation function (\ref{correlator-to-compute}), we insert these wave functions in (\ref{overlapformula}) with 
\be
\delta = 8\pi G_N\, p_1^u p_2^v\,e^{-|x-x'|}.
\ee
Let us study the purely right-sided correlator $\langle V(t_1)W(t_2)V(t_3)W(t_4)\rangle$, with times $t_1 = i\epsilon_1$, $t_2 = t + i\epsilon_2$, $t_3 = i\epsilon_3$, and $t_4 = t + i\epsilon_4$. After some manipulation, the overlap integral becomes
\begin{align}
\frac{C^2a_0^4}{(4\pi)^2}  \int_0^\infty dp\,dq\int_{-\infty}^\infty dx dx' p^{2\Delta_v-1}q^{2\Delta_w-1}e^{-p\cosh (x-x_1)}e^{-q\cosh (x'-x_2)}e^{2\pi i G_N e^{t - |x-x'|}pq/\epsilon_{13}\epsilon^*_{24}},
\end{align}
where
\be
\epsilon_{ij} = i(e^{i\epsilon_i}-e^{i\epsilon_j}),\hspace{20pt}
C = \frac{\pi^2 c_Vc_W}{\Gamma(\Delta_v)\Gamma(\Delta_w)}\left(\frac{1}{2\sin\frac{\epsilon_3-\epsilon_1}{2}}\right)^{\Delta_v}\left(\frac{1}{2\sin\frac{\epsilon_4-\epsilon_2}{2}}\right)^{\Delta_w}.
\ee
Although we used the notation $\epsilon_j$ for the imaginary time parameters, we do not assume that they are small. In particular, by subtracting $\beta/2$ from $\epsilon_1$ and/or adding it to $\epsilon_4$, we can obtain two-sided correlators from this one-sided expectation value.

Without the phase factor $e^{i\delta}$ inserted, the integral would simply give the disconnected overlap, $\langle V V\rangle \langle WW\rangle$. When we introduce the phase factor, the important region of integration is the region in which $\delta \lesssim 1$.\footnote{Here and below, when we say $\delta \lesssim 1$, we mean order one as a function of $t$. If the dimensions of the operators are large, the important region of integration will involve $\delta$ of order $\Delta$.} If the dimensions $\Delta_w,\Delta_v$ are large, one can analyze this integral by saddle point, but we have not found a particularly simple form. However, if we assume that one of the dimensions (say, $\Delta_w$) is much larger than the other, the integral is dominated by the region in which $x'\approx x_2$ and $q\approx 2\Delta_w$. The remaining integral over $p,x$ is a wave function overlap in the background of a fixed shock sourced by the $W$ operator. The $p$ integral can be done exactly, and the $x$ integral can be done by saddle point at large $\Delta_v$. The result is
\be
\frac{\langle V(i\epsilon_1)W(t + i\epsilon_2)V(i\epsilon_3)W(t +i\epsilon_4)\rangle}{\langle V(i\epsilon_1)V(i\epsilon_3)\rangle\,\langle W(i\epsilon_2)W(i\epsilon_4)\rangle} = \left(\frac{1}{1 - \frac{8\pi i G_N \Delta_w}{\epsilon_{13}\epsilon^*_{24}}e^{t-|x_1-x_2|}}\right)^{\Delta_v}.\label{exampleinterp}
\ee

This formula is very similar to the one derived in \cite{SS} for the case of spherically symmetric shocks in a BTZ background. It has also been derived from the large $c$ identity Virasoro block, in \cite{virasoro}. We will make two comments about the formula. First, in order to compare more directly with \cite{SS}, we should move one of the $V$ operators to the left side. We do this by setting $\epsilon_1 = \beta/2=\pi$ and $\epsilon_3 = 0$. This has the effect $\epsilon_{13}\rightarrow -2i$. The behavior is still singular as $\epsilon_{24}\rightarrow 0$, reflecting high-frequency components of the $W$ operator. One way to treat this is to smear the operators in Lorentzian time before taking the $\epsilon$'s to zero. A simpler alternative is to retain some finite but small imaginary time, $\epsilon_4 = -\epsilon_2 = \tau$. This replaces $\epsilon_{24}\rightarrow 2\sin\tau$.

We will emphasize one other point. At large $t$, the correlation function in \cite{SS} decreased as $e^{-2\Delta t}$. Here we find the slower decay $e^{-\Delta t}$. This reflects an interesting difference between localized and spherical shocks. At large $t$ in the localized case that we study here, the important region of $x$ integration is far from the source. In other words, most of the correlation is coming from the tails of the wave functions where the $W$ and $V$ particles are are very far from each other. The slower decay $e^{-\Delta t}$ results from a compromise between the smallness of the transverse tails and the strength of the shock. If the shock is homogeneous, or the space is compact, such compromise is impossible.

\subsubsection*{General background}
On a more general background, we do not know the bulk-to-boundary propagators, so we are not able to give an analysis at the same level of precision. Instead, we will discuss the correlation function in two limits. First, we will identify the place, as a function of $t$, at which gravitational effects start to become important, and second, we will indicate the asymptotic behavior at large $t$ for the case of a compact horizon.

The correlation function starts to be $O(1)$ affected at the point where $\delta(s,b)\sim 1$ when evaluated on the characteristic momenta and transverse positions in the overlap integral {\it without} the $e^{i\delta}$ factor present. In AdS units, the characteristic value of $s$ is of order the relative boost $e^{\frac{2\pi}{\beta}t}$, and the characteristic $b$ is the separation $|x_1-x_2|$. Up to a power law correction, at large $b$ we have $f(b)\sim e^{-\mu b}$, so the correlation function starts to decrease when
\be
G_N e^{\frac{2\pi}{\beta}t}e^{-\mu b} \sim 1
\ee
in AdS units. This translates to 
\begin{align}
t = t_* + \frac{1}{v_B}|x_1-x_2|, \hspace{20pt} t_* = \frac{\beta}{2\pi}\log \frac{1}{G_N},\hspace{20pt} v_B = \sqrt{\frac{d}{2(d-1)}}
\end{align}
here $t_*$ is the scrambling time, and $v_B = \sqrt{d/2(d-1)}$ is the ``butterfly effect velocity'' of \cite{SS,RSS}.\footnote{In a more general black hole background, and in units where the boundary speed of light is one, we have $v_B = \sqrt{2\pi \ell_{AdS}^2/(d-1)r_0\beta}$.} This is in agreement with the analysis in \cite{RSS}. As also emphasized in \cite{kitaev}, we can expand in $G_N$ to see that the initial behavior of the correlation function will be exponential, $\sim e^{\frac{2\pi}{\beta}t}$, independently of the choice of operators $W,V$.

The other regime where we have some control is the large $t$ region on a compact space where we can ignore, at late time, the $x$ dependence of the wave functions. The essential physics here is the longitudinal tails of the wave functions. The important region of integration is where the eikonal phase $\delta$ is of order one, and as $t$ increases, this region moves farther out into the low momentum tails. The behavior of these tails is determined by quasinormal modes. In particular, if the space is compact, the lowest quasinormal frequency dominates, and the wave function is proportional to 
\be
\psi_2(p^v,x) = \int du\,e^{ia_0p^vu/2}\,\langle \phi(u,v,x) \mathcal{O}(t_2,x_2)\rangle|_{v = 0} \propto (p^v)^{\alpha-1}e^{-\frac{2\pi \alpha}{\beta}t_2}
\ee 
Here we have used the fact that the correlation function $\langle \phi \mathcal{O}(t)\rangle$ should decay as $e^{-i\omega t}$ for large $t$, where $Im(\omega)<0$. The constant $\alpha$ is defined by $2\pi \alpha/\beta = i\omega$. The $p^v$ dependence follows from boost covariance.

In thinking about the correlation function, it is very helpful to represent each of the wave functions in unboosted frames, i.e. write all wave functions with boundary times $t = 0$. We include the relative boost of these frames by replacing $s\rightarrow s\,e^{\frac{2\pi}{\beta}t}$ in the overlap integral. For the moment, let us suppress the $x$ dependence of the wave functions and the transverse function $f$. The correlation function is then given by an integral of the form
\be
\langle V\,W(t)V\,W(t)\rangle \propto \int_0^\infty dp\, dq \ p\psi_1(p)^*\psi_3(p) \ q\psi_2(q)^*\psi_4(q) \ e^{iG_N\, pq\, e^{2\pi t/\beta}}\label{generalint}
\ee
At small momentum, these wave functions behave as $\psi_{1,3}\propto p^{\alpha_v-1}$ and $\psi_{2,4}\propto q^{\alpha_w-1}$, using the argument above. At larger momenta, this power law growth is cut off. At large $G_N e^{\frac{2\pi}{\beta}t}$, the important region of integration corresponds to very small $pq$, and correspondingly small values of the wave functions. Let us suppose $\text{Re}(\alpha_w) > \text{Re}(\alpha_v)$. Then the integral prefers to make $p$ small than $q$; the important region of integration will be $q\sim 1$ and $p$ small, and the result will be proportional to $e^{-2i\omega_v(t-t_*)}$. In other words, the late time behavior is determined by the smaller quasinormal frequency.\footnote{In the compact case, there will also be a non-decaying but $1/N^2$ suppressed contribution to the correlation function from the small change in the temperature that results from the application of the operators.}

Let us now put the $x$ dependence back. As the boost increases, the important region of $x,x'$ integration moves to larger and larger $|x-x'|$. The payoff here is that the transverse function $f(|x-x'|)$ decreases exponentially with $x$; the cost is the smallness of the wave function tails. To study this compromise in detail would require some knowledge of the bulk to boundary propagators. We will not attempt this. However, if the space is compact, eventually the important region of integration will consist of antipodal $x,x'$, with one of the coordinates near the $W$ operators. The correlation function will be proportional to $e^{-2i\omega_v(t - t_* - R/v_B)}$, where $R$ is the diameter of the compact space.

We emphasize that the rate of decay in this late time regime is non-universal: it depends on the particular operators through their quasinormal frequencies. This is because the correlation function is dominated by the physics of wave function tails. By contrast, the initial effect on the correlation (discussed above) involved universal exponential behavior $e^{\frac{2\pi}{\beta}t}$. The relevant physics there is the growth of gravitational scattering as a function of energy. Eq.~\ref{exampleinterp} gives an example of how these behaviors can interpolate.

\section{Tree level stringy corrections}\label{stringy}
In this section, we will take a second pass at the amplitude $e^{i\delta(s,b)}$, by studying tree level stringy corrections to the gravity analysis. Our starting point is to assume the formal existence of a worldsheet theory on the black hole background, so that the tree-level contribution to a gauge theory four point function (\ref{WVWV1}), (\ref{WVWV2}), or (\ref{WVWV3}), is given by the world sheet expectation value
\be
\int d^2w\langle V_4(0,0)V_2(w,\bar{w})V_3(1,1)V_1(\infty,\infty)\rangle.\label{tree}
\ee
Here, the vertex operators are determined by the choice of gauge theory correlation function. We will focus on the case of closed string tachyons for simplicity.\footnote{The universality of the Pomeron analysis below was discussed in \cite{Brower:2006ea}.}

The basic parameters of the string theory are the string length $\ell_s = 1/m_s$ and the string coupling $g_s$.   Then $G_N \sim g_s^2 \ell_s^{D-2}$.  To connect to the boundary field theory we will often use the example of N=4 SU(N) SYM where the `t Hooft coupling $\lambda = g_{\rm YM}^2 N =(\ell_{AdS}/\ell_s)^4$ and $g_s = \lambda/N$. 

Because of the kinematics discussed in section \ref{skinematics}, we are interested in a Regge limit of (\ref{tree}) at small $\ell_s/\ell_{AdS}$. In analyzing this amplitude, we will closely follow the work of Brower, Polchinski, Strassler, and Tan \cite{Brower:2006ea}, who studied a similar problem in pure AdS${}_5$. Small $\ell_s/\ell_{AdS}$ allows us to evaluate vertex operators and operator products at Gaussian order. However, as emphasized in \cite{Brower:2006ea}, large $s$ means a large hierarchy of scales in the worldsheet expectation value, so we must retain $\ell_s/\ell_{AdS}$-suppressed corrections to worldsheet conformal dimensions.  We will proceed somewhat formally here, assuming that certain analyticity properties used in \cite{Brower:2006ea} continue to hold in the black hole situation.   The controlled size of the corrections we find and their intuitively plausible form are a self consistency check on this assumption.\footnote{We argue that one possible complicating effect is small in the following section on longitudinal spreading.}

At Gaussian order in the worldsheet theory, the vertex operators are solutions to the wave equation on the black hole background. In the case of interest, these solutions are highly boosted bulk-to-boundary propagators. When the relative boost is large, the zero mode integral in the expectation value (\ref{tree}) will be concentrated near the interaction region near $u = v = 0$. This part of the black hole geometry is weakly curved, justifying a perturbative expansion.   As in the field theory analysis above, it is convenient to decompose these vertex operators into states of near-definite longitudinal momentum. In this section, we will also work in transverse momentum space:
\begin{align}
V_2 &\approx g_2(U)e^{-i q_u U}e^{-ik_2\cdot X} \hspace{20pt} V_1 \approx  g_1(V)e^{-ip_vV}e^{-ik_1\cdot X}\\
V_4 &\approx g_4(U)e^{iq_u U}e^{ik_4\cdot X} \hspace{32pt} V_3\approx g_3(V)e^{ip_vV}e^{ik_3\cdot X}.
\end{align}
The functions $g_j$ are envelope functions, supported near $U = 0$ or $V=0$, but with characteristic momenta that are very small compared to $q_u$ and $p_v$. We wrote `$\approx$' because we are ignoring the weak $V$ dependence of $V_{2,4}$ and the $U$ dependence of $V_{1,3}$. This dependence is important for making the operators (1,1), but in Regge kinematics that is its only role. We will suppress  this dependence and input the (1,1) condition by hand.

\subsubsection*{Flat space Pomeron operator}
For large relative boost, the important region of integration in (\ref{tree}) is at small $w \sim 1/s$, so it is natural to consider an OPE expansion of $V_2 V_4$. In the standard OPE, one organizes the sum in powers of $w$. Here, because we are interested in the region where $w\sim 1/s$, we will keep all powers of $ws$. First, let us consider this product in the Gaussian (flat space) theory, following \cite{Brower:2006ea}. Letting $\sim$ denote agreement of leading terms for small $w$ with $ws$ fixed, we have 
\begin{align}
V_4(0) V_2(w,\bar{w})&\sim (w\bar{w})^{-2 + \ell_s^2 k^2/4r_0^2} \ g_2(U)g_4(U)\,e^{ik\cdot X-iq_uU(w,\bar{w})+iq_u U}\label{pomeron2} \\
&\sim (w\bar{w})^{-2 + \ell_s^2 k^2/4r_0^2} \ g_2(U)g_4(U)\,e^{ik\cdot X-iq_u(w\partial + \bar{w}\bar{\partial})U|_{(0)}}.
\end{align}
Fields without arguments are assumed to be at the origin, and $k = k_4-k_2$. In the first line, we have used that the product of the $U$-dependent factors is nonsingular at $w = 0$, and that the exact vertex operators $V_2,V_4$ are (1,1). In the second line, we retained all powers of $q_uw\partial U$, because after contracting with $V_3$, this will be $\sim ws$. However, we dropped e.g. $q_u w^2\partial^2 U$.

We now integrate this operator $d^2w$, using a formula from \cite{Brower:2006ea}:
\be
\int d^2w\,(w\bar{w})^{-2 + \ell_s^2 k^2/4r_0^2} \ e^{-iq_u(w\partial + \bar{w}\bar{\partial})U} =  \Pi(k^2)\,\left(q_u\partial U q_u\bar{\partial}U\right)^{1-\ell_s^2 k^2/4r_0^2}
\ee
where
\be
\Pi(k^2) = 2\pi\frac{\Gamma(-1+\ell_s^2k^2/4r_0^2)}{\Gamma(2-\ell_s^2k^2/4)}e^{-i\pi+i\pi \ell_s^2k^2/4r_0^2}.
\ee
The resulting ``Pomeron'' operator is
\be
\int d^2w V_4(0)V_2(w,\bar{w}) \sim \Pi(k^2)\,g_2(U)g_4(U)e^{ik\cdot X}\left(q_u\partial U q_u\bar{\partial}U\right)^{1-\ell_s^2 k^2/4r_0^2}. 
\ee
Inserted in (\ref{tree}), it leads directly to standard flat space Regge behavior.

\subsubsection*{First curvature correction}
In a weakly curved background, the above analysis will receive three types of corrections: to vertex operators, to OPE coefficients, and to worldsheet conformal dimensions of the operators appearing in the OPE. Only the latter lead to corrections that grow with center of mass energy. We therefore continue to use OPE coefficients and vertex operators derived in the Gaussian theory. To allow corrections due to shifted conformal dimensions, we rewrite (\ref{pomeron2}) as
\be
V_4(0) V_2(w,\bar{w})\sim w^{L_0-2}\bar{w}^{\bar{L}_0-2} \ g_2(U)g_4(U)\,e^{ik\cdot X-iq_u(\partial + \bar{\partial})U}\label{three}
\ee
in terms of the worldsheet $L_0, \bar{L}_0$ operators. We will work out the correction to $L_0$ at order $\ell_s^2/\ell_{AdS}^2$ using a trick from \cite{Brower:2006ea}. Eq.~(\ref{three}) can be expanded in powers of $\partial U$ and $\bar{\partial}U$. The only terms that survive the integral over $w,\bar{w}$ are diagonal, 
\be
g_2(U)g_4(U)\left(\partial U \bar{\partial}U\right)^{j/2}e^{ik\cdot X}.
\ee
To order $\ell_s^2/\ell_{AdS}^2$, the dimension of such an operator is \footnote{ In the perturbative expansion $c(j)$ takes the form of  a spin $j$ Laplacian plus correction term  acting on the zero mode wavefunctions. This has meaning  analytically continued to complex $j$  \cite{Brower:2006ea}.  We will also assume such analyticity here.
The resulting expression is consistent with the leading term in the $\ell_s/\ell_{AdS}$ expansion of the exact formula $L_0(j,k) = \frac{j}{2} + \frac{k^2 + 1}{4(\ell_{AdS}^2/\ell_s^2-2)}$ for the dimension of such an operator in the NS-NS AdS${}_3$ background of Refs. \cite{Giveon:1998ns,deBoer:1998pp,Kutasov:1999xu,Maldacena:2000hw,Maldacena:2001km}.}
\be
L_0(j,k) = \frac{j}{2} + \frac{\ell_s^2[k^2+c(j)]}{4r_0^2}. \label{pomeron}
\ee
For transverse momenta less than the string scale, $k^2 \ell_s^2 \ll 1$, the flat-space Pomeron operator has spin near two, so to order $\ell_s^2/\ell_{AdS}^2$, we can approximate $c(j)$ by $c(2)$. We determine $c(2)$ by matching the on shell condition $L_0 = 1$ to Einstein's equations for a metric component $h_{uu} \sim \delta(u) e^{i k x}$. This can be read off from Eq.~(\ref{transverseprofile}): $c(2) = \mu^2$. Integrating over $w,\bar{w}$, we find the Pomeron operator
\be
 \Pi(k^2+\mu^2)\delta(U)e^{ik\cdot X}\left(q_u\partial U q_u\bar{\partial}U\right)^{1-\ell_s^2(k^2+\mu^2)/4r_0^2}
\ee\label{pomop}
In this expression, and in the previous paragraph, we are approximating $g_2(U)g_4(U)$ as a $\delta$ function. This is valid in expectation values at high relative boost, since the $V_1,V_3$ vertex operators are almost independent of $U$, and only the integral matters.

\subsubsection*{The amplitude}
We will now use this Pomeron operator to evaluate the amplitude $\delta(s,b)$. The function $\Pi(k^2 + \mu^2)$ has a graviton pole at $k^2 = -\mu^2$, and for $\ell_s^2(k^2+\mu^2)\ll 1$, we can approximate the $\Gamma$ functions by keeping only this pole. Without the factor $(k^2+\mu^2)$ in the exponential and the phase factor from $\Pi$, we would have exactly the vertex operator for the shock wave mode of the graviton. The $\partial U$ and $\bar{\partial}U$ will give factors propotional to $q_u$ when contracted with the vertex operator $V_3$, so the modification from gravity is a factor $(e^{-i\pi/2}\ell_s^2s/4)^{-\ell_s^2(k^2+\mu^2)/2r_0^2}$, where the phase came from $\Pi$. The string-corrected $\delta$ is therefore
\be
\delta(s,b) = \frac{4\pi G_N\,s}{\ell_{AdS}^{D-4}}\int \frac{d^{d-1}k}{(2\pi)^{d-1}} \,\frac{e^{ik\cdot x}}{k^2+\mu^2}\,(e^{-i\pi/2}\ell_s^2s/4)^{-\ell_s^2(k^2+\mu^2)/2r_0^2}.
\ee
where $b = |x|$.

The Fourier transform has two regions of qualitatively different behavior. First, for $b \gg \mu \ell_s^2\log(\ell_s^2s)/r_0^2$, the pole in $k$ dominates the Fourier transform. At the pole, we have $(\ell_s^2s)^{-(k^2 +\mu^2)}=1$, so the answer reduces to the Einstein gravity formula (\ref{gravity}). We reproduce this here, including the large $b$ behavior of the function $f(b)$:
\be
\delta(s,b) =  \frac{4\pi G_N}{r_0^{d-3}}\, s \,\frac{\mu^{\frac{d-4}{2}}e^{-\mu b}}{2(2\pi b)^{\frac{d-2}{2}}}.\label{g}
\ee

Second, for $b\ll \mu \ell_s^2\log(\ell_s^2s)/r_0^2$, the integral is controlled by a saddle point at small imaginary $k$. We find
\be
\delta(s,b) = \frac{4\pi G_N}{r_0^{d-3}}\,s \left(\frac{e^{-\frac{i\pi}{2}}\ell_s^2s}{4}\right)^{-\frac{\mu^2\ell_s^2}{2r_0^2}}\,\frac{e^{-\frac{b^2}{2\rho^2}}}{\mu^2(2\pi \rho^2)^{\frac{d-1}{2}}}.\label{s}
\ee
where
\be
\rho^2=\frac{\ell_s^2}{r_0^2}\log(\ell_s^2s).
\ee
There are several comments to make about this formula. First, as a function of $s$, the formula (\ref{s}) grows more slowly than (\ref{g}). In gravity, the tree amplitude grew with time $t$ as $e^{\frac{2\pi}{\beta}t}$ (remember $s\sim \beta^{-2}e^{\frac{2\pi}{\beta}t})$. Here, we are finding the slower growth
\be
e^{\frac{2\pi}{\beta}\left(1 - \frac{\mu^2\ell_s^2}{2r_0^2}\right)t} = e^{\frac{2\pi}{\beta}\left(1 - \frac{d(d-1)\ell_s^2}{4\ell_{AdS}^2}\right)t}.
\ee\label{delay}
On the right hand side, we have inserted the definition of $\mu$ in (\ref{mu2}). This correction in the exponential should be understood as the first term in a power series expansion in $\ell_s^2/\ell_{AdS}^2$. For large time $t$, the numerical strength of the effect will be sensitive to the full power series. We will return to this point in the Discussion. 

Second, as a function of $b$, the stringy behavior is much smoother than in gravity. The gravitational profile decays exponentially with constant $\mu$. The stringy profile is Gaussian, but at large $s$, the curvature $1/\rho^2$ is very small. This is related to the well-studied phenomenon of transverse string spreading. At high energy, strings spread over a transverse scale $\sqrt{\log s}$, the soft behavior as a function of $b$ can be understood as a consequence of the shock wave source being spread out by this effect.

Third, we note the presence of a small imaginary part in (\ref{s}), from the phase $e^{\pi i \mu^2\ell_s^2/2r_0^2}$. In flat space string scattering, this is due to the inelastic production of long strings in the two-to-two scattering amplitdue. We will have more to say about this effect in \S~\ref{stwo}.

Finally, although we have only studied the tree level amplitude, we expect that the eikonal exponentiation $e^{i\delta(s,b)}$ should be accurate for small $G_N$ with $G_N s$ and $b$ held fixed. This was argued in the flat space analysis of \cite{Amati:1987uf,Amati:1988ww}, where $Re(\delta(s,b))$ was interpreted as arising from a `string corrected' shock wave metric profile. 

\subsubsection*{Longitudinal spreading}\label{longitudinal}
Part of our analysis in this section could have been anticipated based on general expectations about $\sqrt{\log{s}}$ spreading of strings at high energy.   The relation of this phenomenon to black hole horizons is discussed in Ref. \cite{Susskind:1993aa}, which also indicates another type of string spreading, in the longitudinal direction. Although we did not find any clear evidence for longitudinal spreading in our Pomeron analysis above, we will attempt to bound the size of the potential effect based on the calculations in \cite{Susskind:1993aa}. In that paper, the spreading was estimated in light cone gauge. The result was (equation (4.12) of \cite{Susskind:1993aa} converted into our notation)\footnote{We have also corrected a typo in (4.12): the factor of $P$ in the numerator should not be present.}
\be
\Delta V \sim \frac{\ell_s^2}{\epsilon}
\ee
where $\epsilon$ is a smearing interval in the $u$ coordinate, which is being used as light cone time. We interpret this as an uncertainty relation $(\Delta u)(\Delta v) \sim \ell_s^2$. We note that curvatures become large at scales $uv\sim \ell_{AdS}^2$, so longitudinal spreading of this magnitude should not substantially affect our weakly-curved analysis.

We should caution the reader, though,  that the nature of longitudinal spreading is still not well understood,\footnote{But see \cite{longitudinal} for some recent progress.} and that this estimate should be regarded as provisional. A substantially different estimate could make the large curvatures near the singularity more important in the process we are examining and hence decrease our control of the calculation.

\section{Inelastic effects}\label{stwo}

So far we have discussed the elastic part of tree level string scattering.  In this section we turn to  inelastic effects.   We will make estimates using flat space formulas, which capture the basic dynamics.   A standard description of such effects is \cite{Amati:1987uf} which will be the basic reference in this section.   We will describe the kinematics of a collision, as above,  by its impact parameter $b$ and center of mass energy  $\sqrt{s}$.  The impact parameter in the bulk is determined by the transverse coordinates of the local operator in the boundary field theory after folding with bulk to boundary wave functions as discussed above.   In this section for consistency with \cite{Amati:1987uf} we will denote the transverse spreading scale discussed above by $b_I^2 \sim \ell_s^2 \log (s \ell_s^2)$.

\subsubsection*{Oscillator excitation}\label{oscex}

Consider the scattering of two unexcited strings, like gravitons, at lowest order.   There is an amplitude for the oscillator modes of the strings to become excited.   For $b \gg b_I$ this can be understood \cite{Giddings:2006vu}  as the tidal effect of the gravitational field sourced by one string acting  on the other string as an extended object of size $\ell_s$.   This gives an amplitude 
proportional to $ \ell_s^2 \partial_x^2 h(x)$.   For the shock wave profile in (\ref{hx}) this produces a ``diffractive excitation" imaginary part to the phase shift, $\delta_{DE}$,  that is suppressed by a factor $(\ell_s \mu/r_0)^2\sim(\ell_s/\ell_{AdS})^2 \sim 1/\sqrt{\lambda}$ relative to the elastic phase shift $\delta_E$. 

For $b < b_I$ the strength of this effect is smaller,  roughly speaking because the gradient of the `string corrected'  shock profile responsible for (\ref{s}) is smaller.   Relative to the elastic deflection,\footnote{We thank Sasha Zhiboedov for emphasizing this.}  $\delta_{DE}/\delta_{E} \sim \ell_s^2/b_I^2$.   In field theoretic terms this suppression is of order $1/\log N^2$

\subsubsection*{Long string creation}\label{longstr}
Another process that occurs at tree level is the s-channel annihilation of two strings into one long string.  Then the long string decays into multiple short strings, giving an imaginary part to the tree amplitude.  For $b > b_I$ this effect is strongly suppressed,  by the factor $\exp(-b^2/b_ I^2)$ relative to the elastic amplitude.   For $ b < b_I$ this effect is suppressed relative to the real part of $\delta$ by a factor of $\ell_s^2\mu^2/r_0^2\sim 1/\sqrt{\lambda}$.  Here the curvature of the bulk is significant.

\subsubsection*{Multiparticle production}\label{multipar}
At higher order in $g_s$ multiple gravitons can be produced, giving an imaginary part to the scattering amplitude.   The first of these processes is described by the  H-diagram discussed in \cite{Amati:1987uf}.  This process is of order $G_N ^3 s^2$ in  AdS units.    At the characteristic energy for scrambling $G_N s \sim 1$ this effect is suppressed by $g_s^2 \sim 1/N^2$.

\subsubsection*{Black hole production}
At extremely high energies  nonperturbative effects will occur, including macroscopic black hole production \cite{Amati:1987uf,Giddings:2007bw}.  For $b\sim 1$, these occur at energies  $G_N^2 s \sim 1$.   At these energies the elastic scattering amplitude is exponentially small.

\subsubsection*{Effect on correlators}\label{effectcorr}
In this subsection, we will argue that inelastic effects lead to smooth cutoffs at large $s$ in integrals like (\ref{overlapformula}) and that such cutoffs do not change the value of the integral significantly.  The essential point is that the integral is dominated by values of $s$ such that the S-matrix elastic phase is order one. In this range of energy, inelastic effects are subleading.  For simplicity, we will focus on an asymmetric case in which the dimension of one of the $W,V$ operators is much larger than the other. Then we have only a single momentum integral to do.

Let us start with the elastic eikonal approximation, where the correlation function is given by an integral (\ref{generalint}). Replacing the $q$ integral by its central value, we have an integral over $p$ of the form (\ref{intcorr})
\be
I = \int_0^{\infty} dp~ p^a e^{i p h}\label{i}.
\ee
Here we have imagined a compact horizon and an  S wave shock wave function for simplicity (this does not affect our conclusions).  This integral is rendered finite by an $i \epsilon$ prescription that amounts to rotating the integration contour slightly into the upper half plane  (choosing the branch cut for general $a$ away from the first quadrant).   Schematically $p h \sim G_N s$. The integral is dominated by $p$ such that $p h \sim G_N s \sim 1$.  
 
 Now consider  inelastic corrections.    The effect due to excited string oscillator production can be modeled by adding a real part to the exponential of the following form
 \be
I_1 = \int_0^{\infty} dp ~p^a e^{i p h -\eta_1 p h}\label{ione}
\ee
where here $\eta_1$ represents the magnitude of the  correction: $\eta_1 \sim  1/\sqrt{\lambda}$ when $b > b_I$ and $\eta_1 \sim 1/\log N^2$ for $b < b_I$ (long string production effects are of the same order of magnitude).  An estimate of the size of the correction can be obtained by expanding the integrand in $\eta_1$.  This expansion is clearly convergent and the leading term shows that $I_1/I \sim 1 + {\cal O}(\epsilon_1)$.  

Multigraviton production can be modeled by the integral

\be
I_2 = \int_0^{\infty} dp~ p^a e^{i p h -\eta_1 p h -\eta_2 (p h)^2}, 
\ee
where $\eta_2 \sim g_s^2$. The size of the $\eta_2$ correction can again  be found by expanding the integrand. Here the expansion is divergent, but standard estimates show that it is asymptotic.  The first order term gives an accurate estimate for $\eta_2$ small and again the multiplicative correction  is $I_2/I_1 \sim 1+ {\cal O}(\eta_2)$.

Macroscopic black hole production will become significant  at momenta $p_{\rm bh} h \sim 1/g_s^2$. The corresponding  decrease in the integrand in the generalization of $I$ may or may not be analytic, so the above estimates do not immediately apply.   We can model these effects  by a sharp cutoff in $I_2$ at $p_{\rm bh}$.
\be
I_3 = \int_0^{p_{\rm bh}} dp ~p^a e^{i p h -\eta_1 p h -\eta_2 (p h)^2} .
\ee Because the magnitude of the integrand has become small along the real $p$ axis at $p \sim p_{\rm bh}$ we can directly estimate $I_3/I_2 \sim 1+{\cal O}(e^{- \eta_1 p_{\rm bh} h}) \sim 1+ {\cal O}(e^{-\eta_1/g_s^2})$.    Of course there will be small power law corrections, not included in this model,  from these additional effects. 

These arguments show that inelastic effects make a parametrically small correction to the correlator calculations discussed in previous sections.  These small corrections contain interesting physics, though, and we discuss ways of diagnosing them in the following section.   In addition  we discuss a multishock configuration where such effects are order one.

\section{Discussion}\label{discussion}

We have seen that  simple correlators that diagnose scrambling are holographically described by bulk high energy collisions with energies of order $G_N s \sim 1$ in AdS units.  At these energies there are significant corrections to Einstein gravity due to perturbative string effects.   The most important one is elastic transverse spreading.   To understand the imprint of these effects on field theoretic quantities let us first review the behavior of correlators (\ref{WVWV1}), (\ref{WVWV2}), and (\ref{WVWV3}) when the $W$ and $V$ operators are localized in the boundary field theory \cite{SS, RSS}.  

For concreteness consider the one sided arrangement  (\ref{WVWV2})
 
\be
D(|x-y|, t) =\langle W_y(t)V_xW_y(t)V_x\rangle.\label{localized}
\ee
where all operators are on the $R$ side and $x, y$ denote points in the field theory.  As discussed above, (\ref{localized}) is closely related to the commutator $C= \langle[W_x(t), V_y]^2\rangle$:  $D$ is small where the commutator is large, both indicating the action of chaos.

\begin{figure}[ht]
\begin{center}
\includegraphics[scale=.4]{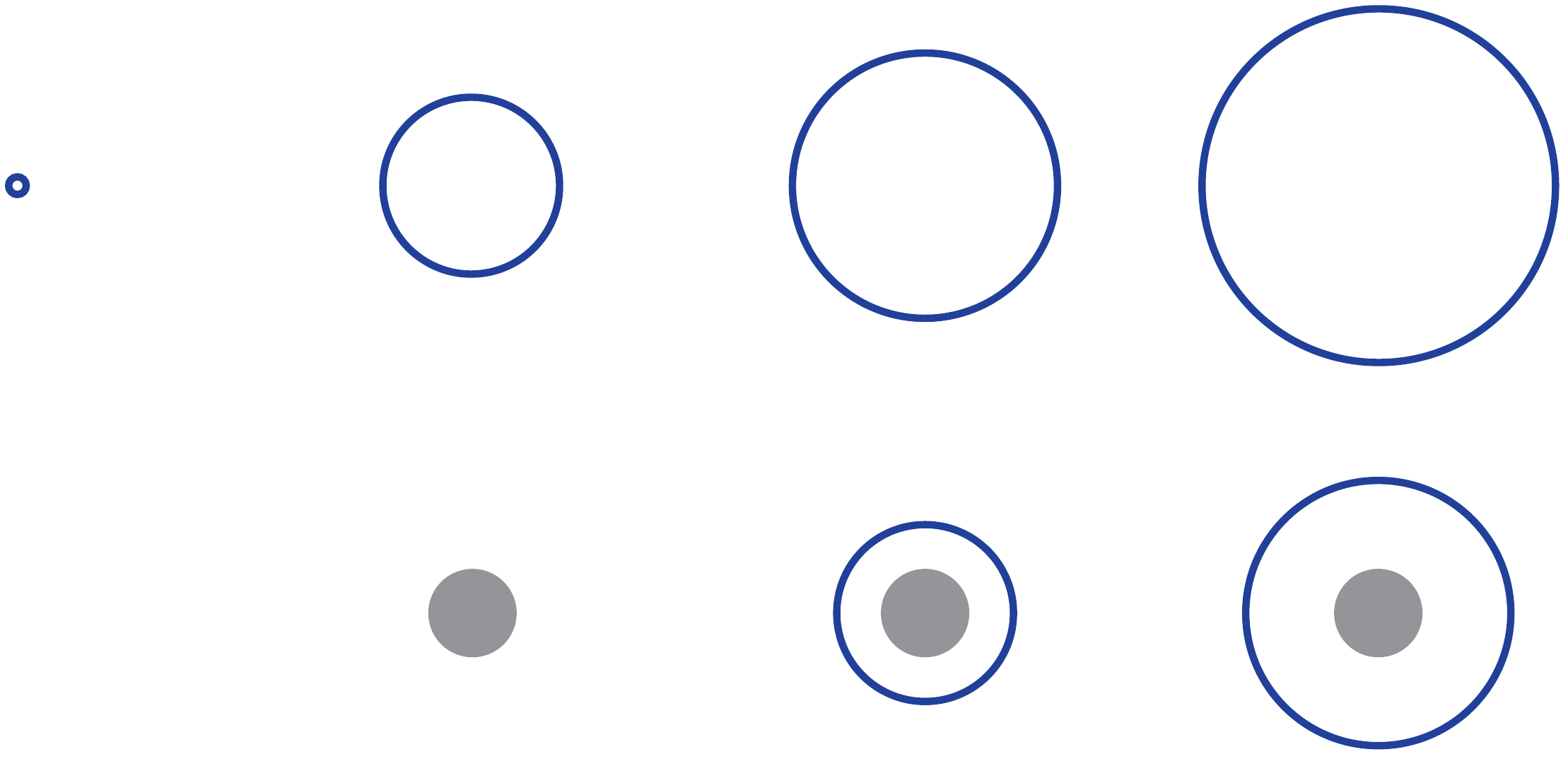}
\end{center}
\caption{First row: ballistic growth of chaotic region in Einstein gravity. \newline Second row:  stringy diffusive disk appears first, then region of ballistic growth moves away faster.}\label{disks}
\end{figure}

The analysis in previous sections shows that  $D$ is determined by folding the scattering amplitude against the wave functions.   The connection between boundary $x$ coordinates and the bulk $x$ profile can be distorted when the scattering is strong and the overlap integral is dominated by tails of the wave functions.   For simplicity we will confine ourselves to regions where the scattering is only moderate and $D$ decreases
 only by an order one relative amount.  In this region we can approximately identify the bulk and boundary transverse coordinate and read off the shock profile from $D(|x-y|, t)$.     The form $f \sim e^{-\mu x}$ multiplied by the boost  $ e^{\frac{2\pi}{\beta}t}$ produces an order one decrease when $|x-y| = v_B(t - t_*)$ where $v_B=2 \pi/(\beta \mu) =\sqrt{d/(2(d-1))}$ is the butterfly effect velocity.   This ballistic growth of chaos is illustrated schematically in the first row of Fig.~\ref{disks} .  The regions inside the circles have $D$ small.  The first observable circle appears at $t_*$. 

But when stringy effects are included we see an interesting modification. From (\ref{g}, \ref{s}) we see that the sharp shock profile is smoothed out by string spreading.   in particular the delay in (\ref{delay}) means that no order one effect appears until a bit after $t_*$, at
\be
t_*^{(\lambda)} = t_* \left(1 + \frac{d(d-1)}{4\sqrt{\lambda}} + ...\right)
\ee  
At $t_*^{(\lambda)}$ the scattering strength for all 
$|x-y|$ less than the spreading scale $\rho^2 = \log(\ell_s^2 s)/(2\sqrt{\lambda})$ is approximately constant because of the smoothing effect of string spreading.  When $t = t_*$, $\rho \sim \frac{1}{\lambda^{1/4}}\sqrt{t_*}
\sim \frac{1}{\lambda^{1/4}}\sqrt{\log{N^2}}~$ so a decreased $D$ appears all across a disk of this parametrically large size. At later times  (\ref{g}, \ref{s}) show that a ballistic circle begins moving away, leaving behind a slowly growing disk dominated by string spreading. The growth is slow when $\lambda$ is large.  A sketch of this is presented in the second row of Fig.~\ref{disks}.  The detailed nature of this growth depends on the convolution of the wave functions with the string amplitude.  We have not be examined this in detail.

In the boundary field theory the $\sqrt{\log s}$ behavior becomes a $\sqrt{t}$ behavior.  This represents a kind of stringy diffusion.   In the bulk this is just a curved space generalization of the branched diffusion on the horizon described in \cite{Mezhlumian:1994pe, Susskind:1993aa}.  In the field theory we see that there is a novel kind of diffusion of chaos occurring in a theory with strings in the dual.   

In \cite{RSS} the connection was made between the ballistic growth of the commutator and the growth of the size of the precursor operator $W_x(t)$.   The ballistic growth with velocity $v_B$ is consistent with the growth of the tensor network describing $W_x(t)$.  Here we see an interesting modification:  stringy diffusion effects smooth out the ballistic growth of the tensor network.  The field theoretic explanation of this, and more generally of the diffusion of chaos is an interesting open problem.  It may be related to the growth of Wilson loop operators in such field theories.

The above analysis applies to large $\lambda$ when the bulk geometry is weakly curved compared to the string scale.   From the field theory perspective it is natural to ask about the behavior of scrambling at small $\lambda$ when the field theory is weakly coupled.   The intuition described in \cite{Sekino:2008he} suggests that  because the strength of  gluon scattering in the gauge theory is of order $\lambda$ at small $\lambda$, the scrambling time should be of order  $\frac{\beta}{\lambda} \log N^2$, parametrically longer than in the gravitational limit.

There is  a potentially  interesting connection between the above and known small $\lambda$ results for high energy scattering in large N SYM.\footnote{The following observations were developed in part in a discussion with Lenny Susskind.}  The parameter $c(2)$ used in (\ref{pomop}) that controls the large $\lambda$ stringy corrections is closely related to the curved space regge intercept $j_0(\lambda)$ that controls high energy AdS scattering, as discussed in ~\cite{Brower:2006ea} .   This parameter can be calculated at small $\lambda$ perturbatively \cite{Lipatov:1976zz,Kuraev:1977fs,Balitsky:1978ic}
 and at all $\lambda$ using integrability techniques.  (Some recent references include \cite{Costa:2012cb, Basso:2014pla,  Alfimov:2014bwa}). Roughly speaking, the eikonal phase at fixed impact parameter behaves like $G_N s^{j_0(\lambda) -1}$.   At large $\lambda$, $j_0(\lambda) = 2 - \frac{c_1}{\lambda^{1/2}}$ giving the standard $G_N s$ gravity limit.\footnote{Here $c_1$, and $c_2$ in the following, are known positive constants.}  At small $\lambda$, $j_0(\lambda) = 1 +c_2 \lambda$, showing that the eikonal phase behaves like $s^{c_2 \lambda}$, increasing very slowly with $s$.   

If we repeated our scrambling analysis in AdS-Rindler coordinates in $D=5$ the eikonal phase would precisely be that of high energy scattering in AdS determined by $j_0$ with $s = e^{\frac{2 \pi }{\beta}t}$ with $\frac{\beta}{2 \pi}= \ell_{AdS}$ .  The correlators would have the schematic magnitude
\be
1 - \frac{const}{N^2}e^{\frac{(j_0(\lambda)-1) 2 \pi }{\beta}t} + {\cal O}(N^{-4})
\ee
giving a scrambling time
\be
 t_* \sim \frac{\beta}{2 \pi (j_0(\lambda)-1)} \log N^2  \sim \frac{ \beta}{2\pi c_2 \lambda} \log N^2
\ee 
as suggested above.

The Regge intercept can be computed perturbatively at small $\lambda$ by resumming a set of gluon ladder diagrams (see the discussion in \cite{banksfestuccia}), with the  diagram with $k$ rungs behaving like  $ (\lambda \log s)^k$.  In AdS-Rindler coordinates this UV divergence becomes a  $(\lambda t)^k$ long time IR divergence for a field theory on hyperbolic space.  This suggests that the weak coupling scrambling time in thermal field theory could be studied by computing IR divergent gluon exchanges perturbatively.

We now turn to inelastic effects. Although their influence on the correlators considered in this paper is parametrically small, as shown in the previous section,  it is still interesting to consider the interpretation of these effects in the scrambling system.  Let us first look at tidal excitation of oscillator modes in the bulk.   Suppose $X$ is a boundary operator dual to a bulk excited string state and consider a correlator of the form 
\begin{equation}
\langle W_L(t) X_L V_R W_R(t) \rangle.
\end{equation}
When $t$ is of order one this correlator is small because $\langle X_L V_R \rangle$ is small, of order $e^{- \lambda^{1/4}}$.  But as $t$ increases tidal excitation increases this overlap, until at $t \sim t_*$ scrambling causes it to exponentially drop.

Long string production and subsequent decay into multiple smaller strings could potentially be diagnosed by correlators of the form 
\begin{equation}
\langle W_L(t) X_L Y_L V_R W_R(t) \rangle
\end{equation}
but because there are many possible final states the amplitude to go into any particular one will be small. Another diagnostic is the initial behavior of the square of the commutator $\langle [V,W(t)]^2\rangle$. In Einstein gravity, the expectation value of the squared commutator starts out at order $N^{-4} e^{\frac{4\pi}{\beta} t}$. This is related to the fact that the tree-level contribution to $\langle V W(t) V W(t)\rangle$ is pure imaginary. However, in string theory, this commutator starts out at order $N^{-2} \lambda^{-1/2} e^{\frac{2\pi}{\beta} t}$. This is because the commutator $[W(t), V]$ has a $N^{-1}\lambda^{-1/4}$ piece corresponding to a long string, in addition to the $N^{-2}$ piece that corresponds to the scattered quanta.

These simple correlators see just  the ``tip of the iceberg"  of scrambling.  All the inelastic scattering effects present in the bulk S matrix are present in the full system even though they make a small contribution to such simple quantities.   The field theoretic objects necessary to fully diagnose them will be very complicated, and will be closely related to those necessary to describe bulk physics behind the horizon.

\begin{figure}[ht]
\begin{center}
\includegraphics[scale=.6]{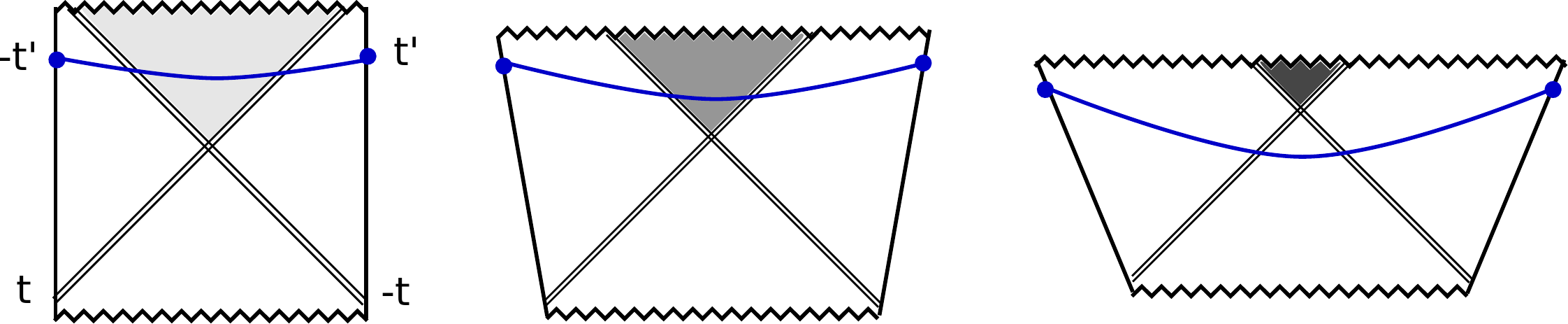}
\end{center}
\caption{Geometries sourced by early shocks on both sides \cite{Shenker:2013yza} in which the collision energy is $G_N^2s\ll 1$ (left) $G_N^2s\sim 1$ (center) and $G_N^2s \gg 1$ (right).}\label{twoshocks}
\end{figure}

In closing we point out one example of a simple correlator involving two shocks where strong inelastic effects do produce a dominant effect. Consider a two-sided correlation function of $V$ operators in a two-shock state
\be
\langle \psi|V_L(-t') V_R(t)|\psi\rangle, \hspace{20pt} |\psi\rangle =  W_L(t)W_R(-t)|TFD\rangle,\label{twocorr}
\ee
where the locations of the times are illustrated in Fig.~\ref{twoshocks}.
 If $\Delta_W\gg \Delta_V\gg 1$, then we can assess the correlation function using geodesics in the background sourced by the $W$ operators. In $D = 3$ spacetime bulk dimensions, these geodesics are real. If the center of mass energy in the collision of the $W$ operators, $s\sim e^{\frac{4\pi}{\beta}t}$ satisfies $G_N^2s \ll 1$, then the geodesic will pass through the post-collision region, but nonlinear effects in the collision will be suppressed (left panel).\footnote{The nonlinear effects due to the purely gravitational interaction will be rather mild in three dimensions, but inelasticities due to other bulk fields, e.g. string states, should be dramatic at these energies. In higher dimensions we expect interesting nonlinearities due to gravity, but the analysis in terms of geodesics is more complicated \cite{Fidkowski:2003nf}.} If $G_N^2s \gg 1$, then nonlinear effects will be important, but the post-collision region will be isolated near the singularity (right panel). However, in the intermediate region $G_N^2 s \sim 1$, the geodesic will pass through a nonlinear post-collision region, suggesting that (\ref{twocorr}) might be sensitive to $G_N^2s$ physics at an order-one level.

\section*{Acknowledgements}
We thank Matthew Dodelson, Alexei Kitaev, Juan Maldacena, Joe Polchinski, Dan Roberts, Eva Silverstein, Andy Strominger, Lenny Susskind, Mark Van Raamsdonk, Xi Yin, and Sasha Zhiboedov for discussions. We are grateful to Raghu Mahajan for initial collaboration and Patrice Gelband for Figure 7. This  work is supported in part by the National Science Foundation under Grants PHY-1316699 and PHY-1314311/Dirac, and by a grant from the John Templeton Foundation. The opinions expressed in this publication are those of the authors and do not necessarily reflect the views of the John Templeton Foundation.  We acknowledge the hospitality of the Aspen Center for Physics and support under its  NSF Grant PHYS-1066293.

\begin{appendix}

\section{Brownian circuit}\label{brownian}
In this appendix, we will compute the correlation function (\ref{WVWV2}) in a quantum system with a random time-dependent Hamiltonian. This system can be understood as a continuum limit of a random circuit in which the couplings change at each time step. This large degree of randomness reduces the computation of the correlator to the solution of a system of $2n$ linear ODEs, where $n$ is the number of sites. The interactions couple all degrees of freedom together, in a way vaguely similar to the dynamics at a single lattice site of a large $N$ gauge theory. We will emphasize three results:
\begin{itemize}
\item The correlator has exponential tails at large $t$.
\item The correlator is very well fit by the holographic shock-wave computation \cite{SS}, $D(t)\approx\left(\frac{1 + a}{1 + a e^{bt}}\right)^c$, with $a\sim 1/n, b\sim 1, c\sim 1$.
\item The time $t_* \propto \log n$ it takes for the correlator to be $O(1)$ affected agrees with an information-theoretic definition of the scrambling time.
\end{itemize}

The specific system that we will work with is the two-body nonlocal Brownian circuit on $n$ spins, studied in section three of \cite{Lashkari:2011yi}. We will follow the conventions from that paper. Roughly, at each time step, the Hamiltonian is proportional to
\be
\sum_{i<j}\sum_{\alpha_i,\alpha_j}J^{(i,j)}_{\alpha_i,\alpha_j}\sigma_{\alpha_i}^{(i)}\sigma_{\alpha_j}^{(j)}
\ee
with Gaussian random $J$ couplings for each pair of sites $i,j=1,...,n$ and pair of Pauli indices $\alpha_i,\alpha_j=0,1,2,3$.

More precisely, the infinitesimal change in the time evolution operator at time $t$ is
\be
U(t+dt) - U(t) = -\frac{n}{2} U(t) dt - i\sqrt{\frac{n}{8(n-1)}}\sum_{i<j}\sum_{\alpha_i,\alpha_j}\sigma^{(i)}_{\alpha_i}\sigma^{(j)}_{\alpha_j}U(t)dB(t)^{(i,j)}_{\alpha_i,\alpha_j}.\label{eom}
\ee
Here, $dt$ is an ordinary infinitesimal, with $dt dt = 0$. The infinitesimal $dB(t)$ is a random variable that has expectation value zero, and has a square that is equal to a Kronecker delta times $dt$,
\be
dB(t)_{\alpha,\beta}^{(i,j)} dB(t)_{\alpha',\beta'}^{(i',j')} = \delta_{i,i'}\delta_{j,j'}\delta_{\alpha,\alpha'}\delta_{\beta,\beta'}dt.
\label{eom2}
\ee
The product $dB dt$ is also zero. 

These rules define an ensemble for the time evolution operator $U(t)$. We are interested in the average, over this ensemble, of the correlation function (\ref{WVWV2}),
\be
D(t) = \mathbf{E}_{U}\Big(\langle \sigma_z^{(2)}(t) \sigma_z^{(1)}\sigma_z^{(2)}(t)\sigma_z^{(1)}\rangle\Big).
\ee
Since the system does not have a fixed Hamiltonian, it does not have a well-defined thermal density matrix. Instead, we will take expectation values in the infinite temperature state. Explicitly,
\be
D(t) =\mathbf{E}_{U}\Big(2^{-n}tr\{U(t) \sigma_z^{(2)}U(t)^\dagger  \ \sigma_z^{(1)} \ U(t)\sigma_z^{(2)}U(t)^\dagger \ \sigma_z^{(1)}\}\Big).
\ee

Using the invariance of the $U(t)$ ensemble under individual rotations of any of the sites, we can conclude that $D$ would be unchanged if we replaced $\sigma_z^{(1)}$ by $\sigma_x^{(1)}$ or $\sigma_y^{(1)}$. On the other hand, if we replace $\sigma_z^{(1)}$ by $1$, we get $2^{-n}tr \ 1 = 1$. Summing over these different replacements and using $\sum_{\alpha=0}^3 \sigma^{(j)}_\alpha M \sigma^{(j)}_\alpha = 2 (tr_{j}M) \otimes 1_j$, we find
\be
3D(t) + 1 = 2^{-n}\mathbf{E}_{U}\Big(2 \ tr_{L-1} \ \left(tr_{1} U(t) \sigma_z^{(2)}U(t)^\dagger\right)^2\Big).\label{yo}
\ee
Here, we are using the notation $L -1$ to mean all sites but the first.

Substituting in for $U(t + dt)$ using Eq.~(\ref{eom}), one finds that the time derivative of $D(t)$ is not purely a function of $D(t)$. This means that we need to study the time evolution of a larger set of variables, in the hopes of finding a closed set of equations. In fact, the RHS is closely related to the purity of the subsytem $L-1$, and the set of all subsytem purities was shown to satisfy a closed set of ODEs in \cite{Lashkari:2011yi}. We will simply borrow this result. Define for any subsytem $A\subset L$
\be
g_A(t) = \mathbf{E}_{U}\Big(2^{-n}tr_A\left(tr_{A_c} U(t) \sigma^{(2)}_zU(t)^\dagger\right)^2\Big).
\ee
The results of \cite{Lashkari:2011yi} show that these functions satisfy
\be
(n-1)\frac{d}{dt}g_A = 2n_{A_c}\sum_{j\in A} g_{A-j} - 4 n_A n_{A_c}g_A + 2 n_A\sum_{j\in A_c}g_{A+j} - \sum_{j \in A, k \in A_c}g_{A + k - j}.\label{yoyo}
\ee
Here, $n,n_A$, and $n_{A_c}$ refer respectively to the number of sites in $L$, $A$, and $A_c = L-A$. Once we solve this system of ODEs with the appropriate initial conditions, we can recover $D(t)$ via Eq.~(\ref{yo}) as
\be
D(t) = \frac{2g_{L-1}-1}{3}.
\ee

Although there are a large number of different subsystems, the symmetry of the initial conditions and the permutation symmetry of the dynamics means that there are only $\sim 2n$ different functions $g_A$. We will parameterize these as $g_k(t)$ and $g^{(2)}_k(t)$, where the former refers to $g_A$ for a subsystem of size $k$ not containing site two, and the latter refers to $g_A$ for a subsystem of size $k$ that does contain site two. The index for $g_k$ runs from $k=0$ to $k = n-1$, and the index for $g^{(2)}_k$ runs from $k = 1$ to $k = n$. 

Using Eq.~(\ref{yoyo}), we find that $g_k,g_k^{(2)}$ satisfy
\begin{align}
\frac{n-1}{k}\frac{d}{dt}g_k &=  2(n-k)g_{k-1}- (5n-5k-1)g_k-g^{(2)}_k+2(n-k-1)g_{k+1}+2g^{(2)}_{k+1}\\
\frac{n-1}{n-k}\frac{d}{dt}g_k^{(2)} &= 2(k-1)g^{(2)}_{k-1}+2g_{k-1}- (5k-1)g^{(2)}_k-g_k+2kg^{(2)}_{k+1}
\end{align}
together with the initial conditions
\begin{align}
g_k(0) &= 0 \\
g^{(2)}_k(0) &= 2^{n-k}.
\end{align}
After solving this system of $2n$ ODEs, we recover the correlator as 
\be
D(t) = \frac{2g^{(2)}_{n-1}-1}{3}.\label{relate}
\ee

One thing is immediately apparent: the late-time asymptotics of the ensemble average of the correlator will be exponential, with a time constant determined by the eigenvalue gap of the system described above. Individual realizations of the Brownian circuit will have $O(2^{-n})$ fluctuations about this mean behavior.

Another feature is apparent if we plot $D(t)$ for different values of $n$. Nothing much happens to the correlator until $t\sim \log n$, after which point the function exponentially decays. To understand this logarithmic behavior, let us change the equations slightly and set $g_k = 0$ for all time. Define $G(x)$ as $2^{k-n}g^{(2)}_k$, evaluated at $x = k/n$. Then the system of ODEs becomes the following PDE:
\be
\partial_t G = -3(1-x)G - 3x(1-x)\partial_x G + \frac{5}{2n}\partial_x^2G + ...
\ee
where the dots are higher order in $\frac{1}{n}\partial_x$. The initial conditions are $G(x,t=0) = 1$, and the correlator is related to $G(1 - 1/n, t)$. Because of the first term, away from $x = 1$, the function $G$ exponentially decreases in a time of order one. However, the factors of $(1-x)$ slow down the evolution near $x = 1$. Naively, it takes a time of order $n$ for the function at $1 - 1/n$ to change significantly. In fact, this analysis is incorrect: expanding near $x = 1$, the equation becomes $\partial_t G = -\partial_y G$, where $y = \frac{1}{3}\log\frac{1}{1-x}$. The time for the correlator to be affected is proportional to $y(x = 1-1/n) = \frac{1}{3}\log n$. This agrees quantitatively with a numerical analysis of the ODEs with $g_k$ replaced by zero. It suggests a picture of fast scrambling consisting of ballistic wave propagation in a logarithmic coordinate.

\begin{figure}
\begin{center}
\includegraphics[scale = .8]{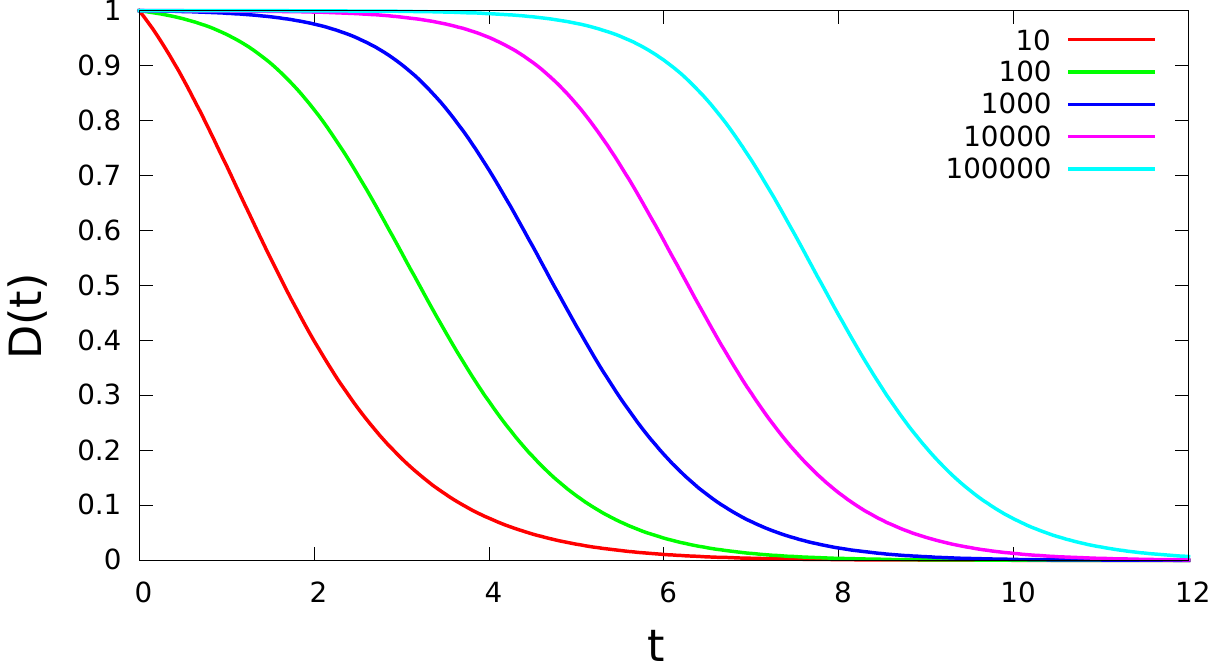}
\caption{The ensemble average of the correlator $D(t)$ for different values of $n$.}
\end{center}
\end{figure}

Finally, let us make contact with an information-theoretic definition of the scrambling time. One definition of scrambling of mixed states proceeds as follows: take a thermal state and perturb one site. How long does it take until this perturbation cannot be detected on any subsystem of size $n-O(1)$? For the Brownian circuit, this time is the same as the time for the correlator to be $O(1)$ affected. We can see this as follows. Take the initial density matrix $\rho = 2^{-n}(1 + \sigma_z^{(1)})$. This is a maximally mixed state on all but the first site. One can show that the purity of subsystem $A$ of size $k$ is equal to $2^{-k} + 2^{-n}g_{A}$. The maximally-mixed value is $2^{-k}$. If the purity is multiplicatively close to this value, then the subsystem cannot be well-distinguished from maximally mixed. The slowest-to-relax purity is the purity of the entire system minus one site, i.e. $g_{L-1}$. The time at which this purity relaxes is related by Eq.~(\ref{relate}) to the time at which the correlator is affected by the perturbation.

\end{appendix}

\end{document}